\documentclass[conference]{IEEEtran}
\IEEEoverridecommandlockouts
\usepackage{cite}
\usepackage{amsmath,amssymb,amsfonts}
\usepackage{algorithmicx}
\usepackage{algorithm}
\usepackage{algpseudocode}
\usepackage{graphicx}
\usepackage{textcomp}
\usepackage{xcolor}
\usepackage{braket}
\usepackage{qcircuit}
\usepackage[hidelinks]{hyperref}
\usepackage{booktabs}
\usepackage[caption=false, font=footnotesize]{subfig}
\ifCLASSOPTIONcompsoc
\usepackage[nocompress]{cite}
\else
\usepackage{cite}
\fi

\def\BibTeX{{\rm B\kern-.05em{\sc i\kern-.025em b}\kern-.08em
T\kern-.1667em\lower.7ex\hbox{E}\kern-.125emX}}
\begin{document}

\title{Accelerating State-Vector Quantum Simulation on Integrated GPUs via Cache Locality Optimization: A Cross-Architecture Evaluation}
\author{
  \IEEEauthorblockN{Gabriel Fernandes Thomaz}
  \IEEEauthorblockA{\textit{Instituto de Pesquisas Eldorado}\\
   Porto Alegre, Brazil \\
  0009-0007-7410-5189}
  \and
  \IEEEauthorblockN{Jerusa Marchi}
  \IEEEauthorblockA{\textit{Instituto de Pesquisas Eldorado}\\
   Porto Alegre, Brazil \\
  0000-0002-4864-3764}
  \and
  \IEEEauthorblockN{Eduarda Rodrigues Monteiro}
  \IEEEauthorblockA{\textit{Instituto de Pesquisas Eldorado}\\
   Porto Alegre, Brazil \\
  0000-0003-2660-3910}
  \and
  \IEEEauthorblockN{Fernando Augusto Caletti de Barros}
  \IEEEauthorblockA{\textit{Instituto de Pesquisas Eldorado}\\
   Porto Alegre, Brazil \\
  0009-0008-5152-7036}
  \and
  \IEEEauthorblockN{Evandro Chagas Ribeiro da Rosa}
  \IEEEauthorblockA{\textit{Instituto de Pesquisas Eldorado}\\
   Porto Alegre, Brazil \\
  0000-0002-8197-9454}
}

\author{
    \IEEEauthorblockN{
        Gabriel Fernandes Thomaz\IEEEauthorrefmark{1},
        Jerusa Marchi\IEEEauthorrefmark{1}\textsuperscript{,}\IEEEauthorrefmark{2},
        Eduarda Rodrigues Monteiro\IEEEauthorrefmark{1}\textsuperscript{,}\IEEEauthorrefmark{3}, \\
        Fernando Augusto Caletti de Barros\IEEEauthorrefmark{1}, \textit{and}
        Evandro Chagas Ribeiro da Rosa\IEEEauthorrefmark{1}\textsuperscript{,}\IEEEauthorrefmark{2}
    }
    \vspace{0.2cm}
    \IEEEauthorblockA{\IEEEauthorrefmark{1}\textit{Instituto de Pesquisas Eldorado}, Porto Alegre, Brazil}
    \IEEEauthorblockA{\IEEEauthorrefmark{2}\textit{Universidade Federal de Santa Catarina}, Florianópolis, Brazil}
    \IEEEauthorblockA{\IEEEauthorrefmark{3}\textit{Pontifícia Universidade Católica do Rio Grande do Sul}, Porto Alegre, Brazil}
}

\maketitle

\begin{abstract}
  The classical simulation of quantum algorithms is a crucial tool for circuit development, testing, and validation. Although acceleration using GPUs significantly reduces simulation time, most high-performance simulators rely on vendor-specific frameworks that target data-center hardware. To broaden access to quantum simulation, this work proposes a vendor-agnostic approach targeting the integrated GPUs commonly found in consumer-grade laptops. A primary challenge in state-vector simulation is its inherently poor spatial locality, which creates a memory bandwidth bottleneck. Consequently, baseline implementations experience a severe degradation in relative GPU speedup as the number of simulated qubits increases. To address this limitation, we introduce a state partitioning optimization that reorganizes the quantum state vector to maximize the last-level cache locality and minimize costly main memory fetches. We evaluate this strategy using a Quantum Phase Estimation algorithm across diverse architectures from Intel, AMD, and Apple. The experimental results demonstrate that the proposed optimization successfully mitigates performance degradation at larger qubit scales. In particular, for a 28-qubit simulation, the optimization reversed a performance deficit on an Intel Core~i5, improving the GPU speedup over the CPU from 0.95$\times$ to 1.89$\times$, and increased the Apple M1 Pro speedup from 3.71$\times$ to 5.88$\times$. Overall, this approach yields consistent execution time improvements, demonstrating the viability of integrated GPUs for efficient quantum circuit simulation.
\end{abstract}

\begin{IEEEkeywords}
  Quantum Circuit Simulation, State-Vector, Integrated GPU, Cache Locality, Multi-Vendor
\end{IEEEkeywords}

\section{Introduction}

Quantum simulation is an important component for quantum computing development toolchain. Simulators enable the execution of quantum algorithms on classical computers and are widely used for algorithm design, testing, and validation~\cite{murillo2025}. Although general quantum simulation requires resources that grow exponentially with the number of qubits~\cite{preskill2018}, simulation remains relevant for research and education purposes. Many algorithms of practical interest can be studied on small scales, where the limited number of qubits does not justify the cost and limited availability of quantum hardware. In addition, access to quantum devices may be constrained by financial cost, queue times, or institutional availability, particularly in developing countries. For small circuits, the waiting time for hardware execution can exceed the time required for classical simulation. Moreover, simulation provides exact results without the effects of noise and decoherence, which facilitates algorithm analysis.

Several approaches exist for the classical simulation of quantum computers~\cite{cicero2026}, including tensor networks, decision diagrams, and sparse-state representations. Under specific restrictions, such as Clifford circuits~\cite{gottesman1997}, polynomial-time simulation is possible using methods such as the stabilizer formalism.
However, no single simulation technique is optimal for all classes of algorithms: some of them demonstrate to be efficient for particular circuits.

Nevertheless, the state vector method seems to be the most direct approach: the quantum state is stored as a complex vector, and quantum gates are applied as linear transformations following the standard linear-algebra formulation of quantum mechanics. This implementation closely reflects the algebraic description of the theory without relying on alternative representations or reductions, providing  predictable time and memory requirements and serving  as a general baseline method.

State vector simulation is dominated by linear algebra operations, primarily structured matrix-vector multiplications. These operations exhibit a high degree of data parallelism, which makes them suitable for multithreaded and GPU execution. Furthermore, the regular structure of the quantum gates allows the matrix operations to be implemented implicitly, avoiding the explicit construction of large matrices~\cite{smelyanskiy2016}. Also, the method requires limited CPU-GPU communication, with the CPU mainly responsible for kernel dispatch.

Most high-performance quantum simulators target data-center GPUs with high memory bandwidth, large core counts, and fast interconnects such as NVLink, enabling multi-GPU simulations of larger systems~\cite{bayraktar2023}. But this infrastructure is expensive and not available to many research groups. Consumer discrete GPUs offer a more accessible alternative and are supported by several existing simulators. An even more widely available platform is the integrated GPU (iGPU) present in most modern laptops and some desktop processors. Integrated GPUs share memory with the CPU and provide fewer compute units, but their ubiquity makes them a relevant target for low-cost quantum simulation.

In this work, we investigate the use of integrated GPUs for state vector simulation~\cite{smelyanskiy2016,guerreschi2020,bayraktar2023}. We evaluate systems based on Intel, AMD, and Apple architectures. Our experiments show that, once kernel launch overhead is amortized, integrated GPUs can provide speedup over CPU execution for moderate circuit sizes. However, this advantage decreases as the number of qubits grows. To address this limitation, we propose an optimization that reorganizes the state vector to improve cache spatial locality, resulting in measurable performance gains.

The remainder of this paper is organized as follows. Section~\ref{sec:related_work} reviews previous work on GPU-based quantum simulation and cache locality optimizations. Section~\ref{sec:state_vector_simulation} describes the state vector algorithm and the proposed state partition strategy. Section~\ref{sec:benchmark_setup} presents the evaluation methodology, metrics, and hardware platforms. Section~\ref{sec:results_and_discussion} analyzes the experimental results, and Section~\ref{sec:conclusion} concludes the paper.

\section{Related Work}\label{sec:related_work}


The landscape of classical quantum circuit simulation is broad and heterogeneous, encompassing multiple algorithmic strategies, hardware, and optimization techniques. To provide a coherent foundation for our contribution, this section reviews the main classes of simulators and acceleration strategies most closely related to our work. We begin with state-vector simulators, the dominant approach for the exact simulation of quantum circuits. Next, we discuss GPU-accelerated frameworks that leverage vendor-specific hardware to reduce execution time in Section~\ref{subse:gpu-accelerated}. Finally, in Section~\ref{subse:cache-locality}, we examine locality-aware optimization techniques that mitigate memory bandwidth bottlenecks, a central issue for state-vector simulation and one directly addressed by our proposed method. Throughout this review, we contrast previous approaches with our vendor-agnostic, cache-optimized strateg for iGPUs.

State vector simulation is one of the most widely adopted methods for accurate emulation of quantum circuits, as it directly represents a quantum state of $n$ qubits using a complex vector of size $2^n$. One of the first high-performance implementations operating under this model is qHiPSTER~\cite{smelyanskiy2016}, later expanded and renamed Intel Quantum Simulator (Intel-QS)~\cite{guerreschi2020}. The qHiPSTER framework introduced a distributed memory formulation for applying quantum gates based on MPI, allowing the complete state vector to be partitioned across computing nodes, with communication patterns optimized to support generic one- and two-qubit gates at scale. Its initial version demonstrated efficient simulation of circuits up to 40 qubits in \emph{High Performance Computing} (HPC) systems, combining multithreading, vectorization, and communication overlap techniques.

Based on this foundation, Intel-QS has evolved into a production-ready simulator focused on cloud and HPC usability. The updated system introduced an execution model ``pool''  that allows the simultaneous simulation of multiple related quantum circuits, significantly improving throughput in variational algorithms and ensemble-based workloads. Current studies demonstrate Intel-QS scaling to more than 40 qubits, benefiting from optimized memory layouts, communication scheduling, and careful exploitation of CPU parallelism.

While these simulators provide robust high-performance baselines, they share a limitation common to most CPU-centric state vector tools: their performance degrades as the qubit count increases due to the inherently poor spatial locality of the state vector data structure. The technologies mentioned rely heavily on multicore CPUs and distributed memory clusters, and while they achieve excellent scalability in HPC infrastructure, they do not address the growing need for vendor-agnostic simulators for consumer hardware, nor do they address the cache locality degradation that suppresses iGPU speed gains in memory-bottlenecked scenarios. This contrasts with the goal of the present work, which targets integrated GPUs and introduces a state partitioning optimization specifically designed to maximize cache reuse and mitigate memory bandwidth limitations. As our evaluation demonstrates, such optimizations allow integrated GPUs to achieve competitive and scalable acceleration on Intel, AMD, and Apple architectures, even in regimes where conventional simulations would suffer from bandwidth-induced performance collapse.

\subsection{GPU-Accelerated Simulators}\label{subse:gpu-accelerated}

GPU acceleration has become a central strategy for reducing execution time in classical quantum circuit simulation. Proof of this is NVIDIA's cuQuantum SDK~\cite{bayraktar2023}, a representative, production-grade toolset that exposes GPU-optimized primitives for state vector (cuStateVec) and tensor network (cuTensorNet) backends. These building blocks are designed for modern NVIDIA architectures and enable substantial speed gains over CPU-only execution, with direct scaling from single GPU to distributed multi-GPU deployments in HPC or cloud environments. In other words, cuQuantum is a vendor-specific acceleration layer that subsequent simulators can adopt to achieve high performance on discrete NVIDIA GPUs.

In addition to libraries, complete simulations have also been designed around GPUs. HyQuas~\cite{zhang2021} proposes a hybrid partitioner that automatically selects the most suitable simulation method for each region of the circuit, combining shared memory optimizations (OShareMem) and matrix multiplication rework (TransMM) that leverages cuBLAS/Tensor cores. HyQuas adds a GPU-centric communication pipeline to scale across multiple GPUs, reporting a speed increase of up to 10.7$\times$ on individual GPUs and 227$\times$ on GPU clusters compared to previous systems, explicitly targeting discrete CUDA devices and NVLink/NCCL interconnects.

Among general-purpose simulations, QuEST stands out for unifying multithreading, distributed MPI, and GPU execution in a single C library. Jones \textit{et al.}~\cite{jones2019} demonstrate strong/weak scalability down to 38 qubits across thousands of CPU cores and document GPU support as part of a high-performance portable stack that runs from laptops to supercomputers, a novelty at the time for bringing together CPU paths, distributed memory, and accelerators into a single interface. Still, the GPU path in QuEST follows the conventional model of discrete accelerators, focusing on HPC environments.

With a specific focus on large-scale performance in heterogeneous clusters, SV-Sim~\cite{li2021} is a PGAS-based state vector simulator with GPU-centric and CPU-GPU-centric communication strategies to reduce choppy data exchange between nodes. It supports NVIDIA and AMD GPUs and integrates with front-ends such as Qiskit, Cirq, Q\#, OpenQASM, and QIR, presenting scalability on the supercomputers such as Summit, Theta, Cori with GPUs NVIDIA V100, NVIDIA A100, and AMD MI100. The emphasis is on distributed execution across multiple GPUs, communication/compute orchestration, and end-to-end HPC throughput.

In contrast to the solutions above, which assume discrete GPUs and vendor-specific stacks, our study targets GPUs integrated into consumer-grade architectures (Intel, AMD, Apple), with a direct focus on the spatial locality of the state-vector. Unlike cuQuantum, HyQuas, QuEST, and SV-Sim, which maximize FLOPs and high-throughput communication on discrete GPUs and/or clusters, we address the bandwidth and cache bottleneck of the state-vector simulation via partition reorganization to maximize last-level cache locality in the iGPU scenario. While the cited works show how to achieve high performance in data-center GPUs and specialized interconnects, our result demonstrates that iGPUs can achieve consistent gains when the state layout is adjusted to cache behavior, filling a gap little explored by the literature.

\subsection{Cache Locality Optimizations}\label{subse:cache-locality}

State-vector simulation is fundamentally constrained by memory-system behavior: as the target-qubit index $t$ increases, the stride-$2^t$ access pattern forces the simulator to read and update amplitudes that are increasingly distant in memory. This quickly exhausts cache capacity and shifts execution from compute-bound to memory-bound, causing performance degradation even on modern multicore CPUs and discrete GPUs. To mitigate this, prior work has explored techniques that improve spatial and temporal locality through (i) circuit-level restructuring, (ii) hierarchical decomposition of the circuit into smaller state vectors, and (iii) architecture-aware tiling for GPU shared memory.

Doi and Horii~\cite{doi2020} introduce a cache-blocking technique in Qiskit Aer, where circuits are transpiled such that gates are reassigned to low-index qubits within fixed-size chunks, and noiseless SWAP gates are inserted when qubits must migrate across chunk boundaries. This remapping ensures that gate operations operate on contiguous sub-states that fit into fast memory (GPU memory or CPU cache), reducing cross-chunk traffic and enabling scalable multi-GPU/MPI execution. The method is circuit-centric: locality is improved by changing the circuit structure itself, rather than altering the state layout.

The hierarchical state-vector simulation model~\cite{fang2022} represents the quantum circuit as a dependency graph and then partitions this graph into acyclic subgraphs, each corresponding to a subcircuit with a limited data footprint. These subcircuits are simulated hierarchically by constructing and evolving smaller state vectors, which reduces the number of full-vector passes and improves temporal locality during gate application. By lowering the working-set size and confining operations to smaller blocks of amplitudes, the method achieves better time-to-solution and improved scalability on multicore systems. Unlike circuit-level cache blocking, this approach enhances locality by graph-based partitioning and hierarchical execution rather than explicit cache-aware data reordering.

HyQuas~\cite{zhang2021} incorporates techniques relevant to locality by selecting between two GPU execution modes: (i) ~OShareMem, which tiles gate operations into on-chip shared memory, enhancing reuse and reducing global memory traffic; and (ii) TransMM, which reformulates gate application as batched GEMMs to exploit Tensor Cores and cuBLAS. While HyQuas does not target CPU cache locality, it is an example of architecture-aware locality optimization, tailored specifically to CUDA-class discrete GPUs~\cite{zhang2021}.

While these approaches improve locality through circuit transformations, graph-based subcircuit extraction, or GPU-specific tiling strategies, they all assume high-bandwidth, discrete hardware environments, either multi-GPU clusters (Qiskit Aer cache blocking), NUMA-aware multicore servers (hierarchical state-vector simulation), or CUDA-optimized accelerators with large on-chip shared memory (HyQuas). None of them directly addresses locality constraints in integrated GPUs, where memory bandwidth is significantly lower and cache resources are shared with CPU cores.

Our experimental results show that cache-aware reorganization successfully mitigates performance degradation at larger qubit scales and restores integrated GPU speedup, where locality-unaware approaches typically collapse. This fills a gap in the prior literature by demonstrating the viability of iGPU-based quantum simulation through cache-centric, rather than circuit-centric, or GPU-specific, locality optimizations.

\section{State Vector Simulation}\label{sec:state_vector_simulation}

In this section, we describe the state vector simulator used in our work. Section~\ref{subse:base_algorithm} presents the base algorithm for parallel execution, using the approach adopted in simulators such as qH$i$PSTER~\cite{smelyanskiy2016} and QuEST~\cite{jones2019}.
Section~\ref{subse:dividing_the_quantum_state} introduces the proposed optimization based on dividing the quantum state into blocks, a technique related to those employed in distributed~\cite{guerreschi2020} and multi-GPU simulations~\cite{bayraktar2023}, here adapted to reduce cache misses on a single processor. Implementation details are presented in Section~\ref{subse:gpu_implementation}, and the impact of the optimization is analyzed in Section~\ref{sec:results_and_discussion}.

\subsection{Base Algorithm}\label{subse:base_algorithm}

In the state vector simulation, the state of an $n$-qubit system is represented by an array of complex numbers with $2^n$ elements.
Although the mathematical formulation associates each computational step with a $2^n \times 2^n$ unitary matrix, a universal simulator does not need to explicitly construct this matrix.
Gate application can be expressed directly in terms of operations on the state vector with two fundamental cases: single-qubit gates and controlled gates.

The application of a single-qubit gate $U$ on qubit $k$ of an $n$-qubit system corresponds to the operator:
\begin{equation}
  U_k = I^{\otimes k} \otimes U \otimes I^{\otimes n-k-1}.
\end{equation}

\begin{figure}[htbp]
  \centering
  \includegraphics[width=.8\linewidth]{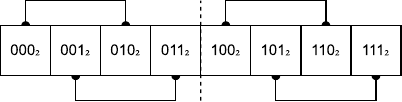}
  \caption{Memory positions read and written when applying a single-qubit gate to qubit index~1 of a three-qubit system.
    The dashed line separates the regions processed by each iteration of the outer loop in Algorithm~\ref{fig:apply_gate}.
  Lines connecting elements represent the pairs accessed in the inner loop.}
  \label{fig:3qubits}
\end{figure}

Algorithm~\ref{fig:apply_gate} shows the procedure for applying $U_k$.
Each iteration of the inner and outer loop is independent and can be executed in parallel.
Within the inner loop, two positions of the state vector are read and written, and the access pattern does not create data races.
Figure~\ref{fig:3qubits} illustrates the memory addresses accessed when a gate is applied to qubit index~1 of a three-qubit system, explicitly the absence of accesses conflicts.

\begin{algorithm}[htbp]
  \centering
  \caption{Single-qubit gate application algorithm.
    $S$: state vector of size $2^n$;
    $U$: $2 \times 2$ single-qubit gate;
  $t$: target qubit index (starting at 0).}
  \begin{algorithmic}[1]
    \Procedure{\textsc{ApplyGate}}{$S$, $U$, $t$}
    \For{$i = 0$ \textbf{to} $2^n-1$ \textbf{step} $2^{t+1}$}
    \For{$j = i$ \textbf{to} $i + 2^{t}-1$}
    \State $a \gets S[j]$ \Comment{State $\ket{0}$ amplitude}
    \State $b \gets S[j + 2^{t}]$ \Comment{State $\ket{1}$ amplitude}
    \State $S[j] \gets U_{1,1} \cdot a + U_{1,2} \cdot b$
    \State $S[j + 2^{t}] \gets U_{2,1} \cdot a + U_{2,2} \cdot b$
    \EndFor
    \EndFor
    \EndProcedure
  \end{algorithmic}
  \label{fig:apply_gate}
\end{algorithm}

Controlled gates can be implemented by conditioning the execution of the inner loop of Algorithm~\ref{fig:apply_gate}.
For a control qubit $c$, the $j$-th iteration is executed only if the $c$-th bit of $j$ is 1.

This algorithm is also the basis for distributed quantum simulation.
The index of the target qubit determines how pairs of amplitudes are accessed, which in turn defines whether operations can be performed locally or require communication.
In the next section, we employ the same observation not for distribution across nodes but to improve cache locality.

\subsection{Dividing the Quantum State}\label{subse:dividing_the_quantum_state}

The proposed optimization divides the quantum state into blocks in order to improve cache locality.
This strategy is similar to that used in distributed and multi-GPU simulators, but here it is evaluated for a single processor.

In Figure~\ref{fig:3qubits}, when a gate is applied to qubit index~1, part of the amplitude pairs accessed lies in the first half of the vector, while the corresponding pairs lie in the second half.
In a distributed setting, each half could be stored on a different node, allowing operations on some qubits to be performed independently.
Operations on higher-index qubits would require communication between nodes.
This observation motivates the division of the state into blocks.

\begin{figure*}[htbp]
  \centering
  \includegraphics[width=.85\linewidth]{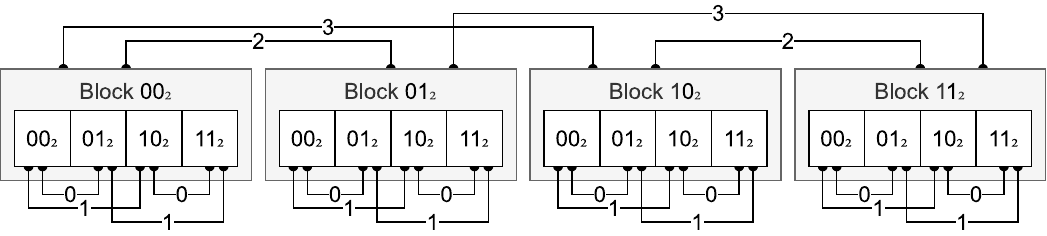}
  \caption{Four-qubit system with the state divided into four blocks.
    Qubits 0 and 1 correspond to local qubits, whereas qubits 2 and 3 correspond to global qubits.
  The connecting lines represent the pairs of amplitudes accessed during the application of a gate to a specific qubit index.}
  \label{fig:24qubits}
\end{figure*}

After partitioning, qubits are classified as \emph{local} or \emph{global}.
Operations on local qubits access data inside a single block, while operations on global qubits require data from multiple blocks.
Figure~\ref{fig:24qubits} illustrates a four-qubit system divided into four blocks.
For a simulation of $n$ qubits divided into blocks of size $2^l$, qubits with index $0$ to $l-1$ are local and those with index $l$ to $n-1$ are global.

When an operation targets a global qubit, a SWAP operation is performed to map that qubit to a local position within the block.
A least-recently-used (LRU) policy selects the local qubit to be swapped.
The simulator maintains a logical-to-physical qubit mapping that is updated after each SWAP.

The optimization relies on selecting a block size that fits in the last-level cache (LLC). To improve the reuse of cached data, operations acting on local qubits are delayed rather than executed immediately. These deferred operations are batched and executed only when an operation targeting a global qubit is required. During this execution phase, all delayed local operations are applied entirely to a single block before the simulator transitions to the next block, effectively keeping the active data resident in the cache. Furthermore, for controlled operations, only the target qubit is considered when determining if a global SWAP is necessary. If a global qubit acts solely as a control, the simulator simply evaluates whether the operation applies to a specific block by checking the block's index, thereby avoiding unnecessary state reorganizations.

Despite of these optimizations, operations targeting global qubits inherently require block reorganizations, which typically generate additional cache misses. Consequently, a trade-off exists: small blocks increase the number of global qubits and required SWAPs, whereas large blocks reduce intra-block locality. This trade-off between block size and cache behavior is evaluated in Section~\ref{sec:results_and_discussion}.

\subsection{GPU Implementation}\label{subse:gpu_implementation}

The algorithm described above was implemented for both CPU and GPU execution.
The CPU version is based on the multithreading KBW simulator of the Ket quantum programming platform~\cite{rosa2026}, implemented in Rust.
The GPU implementation uses CubeCL 0.9, a Rust-based General-Purpose GPU framework, with the graphics library \texttt{wgpu} backend, enabling execution through Vulkan and Metal on Linux, macOS, and Windows.

As detailed in Section~\ref{sec:benchmark_setup}, the implementation was evaluated on Intel, AMD, and Apple processors under Linux and macOS.
The simulator was also validated on Windows systems using the same code base.

\section{Experimental Setup}\label{sec:benchmark_setup}

\begin{table*}
  \centering
  \caption{Hardware specifications of the consumer-grade laptop systems used to evaluate the proposed optimization.}  \label{tab:specs_notebooks}
  \begin{tabular}{lllll}
    \toprule
    {Laptop} & {CPU} & {GPU} & {OS} & {Kernel} \\
    \midrule
    Inspiron 13 5330 & Intel Core i5-1340P & Intel Iris Xe Graphics & Arch Linux  & Linux 6.18.9-zen1-2-zen \\
    Yoga Slim 7 14ILL10 & Intel Core Ultra 7 258V & Intel Arc Graphics 130V & Debian GNU/Linux 13  & Linux 6.12.69+deb13-amd64 \\
    IdeaPad 3 15ALC6 & AMD Ryzen 5 5500U & AMD Lucienne & Fedora Linux 43  & Linux 6.18.7-200.fc43. \\
    MacBook Pro (14-inch, 2021) & Apple M1 Pro (8 cores) & Apple M1 Pro (14 cores) & macOS Sequoia 15.6.1 & Darwin 24.6.0 \\
    \bottomrule
  \end{tabular}
\end{table*}

We evaluated the proposed optimization using a Quantum Phase Estimation algorithm, detailed in Section~\ref{subsec:qfe}. The evaluation considered three metrics: wall-clock time, cache misses, and global-local qubit SWAPs, which are discussed in Section~\ref{subsec:metrics}. The benchmarks were executed on four distinct systems encompassing Intel, AMD, and Apple processors with integrated GPUs; the hardware specifications are presented in Section~\ref{subsec:hardware}. The benchmark results and subsequent analysis are presented in Section~\ref{sec:results_and_discussion}.

\subsection{Quantum Phase Estimation Algorithm}\label{subsec:qfe}

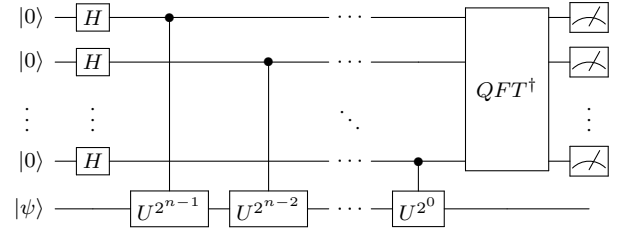
\begin{figure}[htbp]
  {  \footnotesize
    $$
    \begin{array}{c}
      \Qcircuit @C=1em @R=.7em {
        \lstick{\ket{0}}     & \gate{H}      & \ctrl{4}           & \qw                & \qw  &\cdots && \qw            & \multigate{3}{QFT^\dagger} & \meter        \\
        \lstick{\ket{0}}     & \gate{H}      & \qw                & \ctrl{3}           & \qw  &\cdots && \qw            & \ghost{QFT^\dagger}        & \meter        \\
        \lstick{\vdots \ \ } & \push{\vdots} &                    &                    &      &\ddots &&                &                            & \push{\vdots} \\
        \lstick{\ket{0}}     & \gate{H}      & \qw                & \qw                & \qw  &\cdots && \ctrl{1}       & \ghost{QFT^\dagger}        & \meter        \\
        \lstick{\ket{\psi}}  & \qw           & \gate{U^{2^{n-1}}} & \gate{U^{2^{n-2}}} & \qw  &\cdots && \gate{U^{2^0}} & \qw                        & \qw
      }
    \end{array}
    $$
  }
  \caption{Circuit for the Quantum Phase Estimation Algorithm.}\label{fig:qfe}
\end{figure}

Given a unitary operator $U$ and its eigenstate $\ket{\psi}$, the Quantum Phase Estimation (QPE) algorithm~\cite[\S 5.2]{nielsen2010} determines the angle $\theta$ of the corresponding eigenvalue $e^{2\pi i \theta}$. As presented in Figure~\ref{fig:qfe}, the algorithm circuit utilizes the inverse Quantum Fourier Transform ($QFT^\dagger$), which is implemented using controlled-phase and Hadamard gates.

For our benchmark, we simulate $n+1$ qubits, where $n$ qubits control the application of the exponential of $U$, and one qubit is prepared in the state $\ket{1}$ to act as the eigenstate of $U$. For the unitary $U$, we use a phase gate defined as:
\begin{equation}
  U(n) = P(2\pi 2^{-n}) =
  \begin{bmatrix}
    1 & 0 \\ 0 & e^{2\pi i 2^{-n}}
  \end{bmatrix}.
\end{equation}

In this $(n+1)$-qubit simulation, the angle $\theta$ is $2^{-n}$. This specific angle facilitates straightforward verification of the execution, as the measured bitstring is expected to be exactly $00\dots01$. Additionally, the exponential of $U$ can be computed directly as:
\begin{equation}
  U(n)^a = P(2\pi 2^{-n}a).
\end{equation}

The experiment was implemented in Rust. This implementation resulted in a compiled binary that takes the number of qubits and the block size as execution arguments.

\subsection{Metrics}\label{subsec:metrics}

To automate the execution of the experiments, a Python script was used. We measured the wall-clock time required to execute the simulation from program initialization to termination. For cache metrics, the Linux \texttt{perf} tool was used to record cache references and cache misses from the last-level cache (LLC) via the processor's Performance Monitoring Unit (PMU). Because these hardware registers cannot be directly accessed during GPU execution, the CPU results served as a proxy to analyze the cache behavior. In both the CPU and GPU executions, we also recorded the total number of SWAP operations required during the algorithm's execution.

Analyzing cache misses helps to understand whether the proposed block division strategy improves cache locality, which can positively impact execution time. Conversely, the SWAP count indicates the additional operations required to perform the necessary gate applications, which negatively impacts execution time. The working hypothesis is that when the block size fits within the LLC, cache locality improves, reducing wall-clock time. Smaller block sizes, however, are not expected to improve locality as the increased frequency of SWAPs will trigger additional cache misses.

The evaluation varied the number of qubits from 4 to 28, using single-precision floating-point numbers for the simulation. For executions involving 16 or more qubits, the block size was varied between 16 and $n-1$ qubits (where $n$ is the total number of qubits). The baseline for this benchmark is the execution without dividing the quantum state. We swept across different block sizes to identify the optimal configuration.

For the Apple M1 Pro system, cache miss data was not collected due to architectural limitations in reading the PMU registers on macOS. However, cache behavior in this system can be inferred by observing the optimal block size.

Each benchmark configuration was executed three times and the median was recorded. A two-second sleep interval was introduced between executions to allow the processor to cool down. This interval is necessary because thermal throttling measurably impacts execution time, particularly on laptops, which are the primary focus of these experiments.

\subsection{Hardware}\label{subsec:hardware}

We evaluated the proposed optimization on the four systems detailed in Table~\ref{tab:specs_notebooks}. It is important to note that these systems feature processors from different generations and operate with varying thermal design powers (TDPs). Consequently, the purpose of this evaluation is not to conduct a direct performance comparison between hardware. Instead, our objective is to assess the efficacy and behavior of the optimization independently across a diverse set of consumer-grade architectures.

We note that the hardware selection for this study was dictated by the most stringent of resource constraints: the immediate physical availability of the authors' personal computers. Nevertheless, this ensemble of daily drivers effectively demonstrates the optimization's performance across a diverse, if unintentional, variety of hardware ecosystems.

\begin{figure*}[!t]
  \centering

  \subfloat[Core i5-1340P (Baseline).]{
    \includegraphics[width=.23\linewidth]{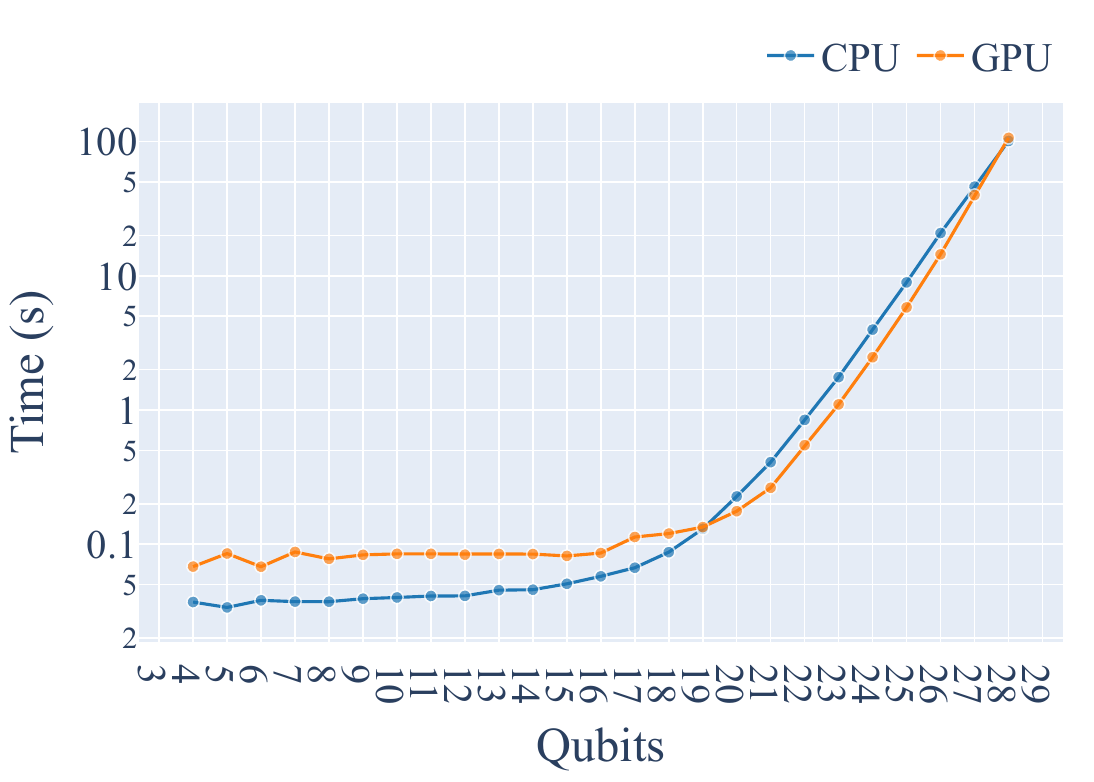}
  }\hfill
  \subfloat[Core Ultra 7 258V (Baseline).]{
    \includegraphics[width=.23\linewidth]{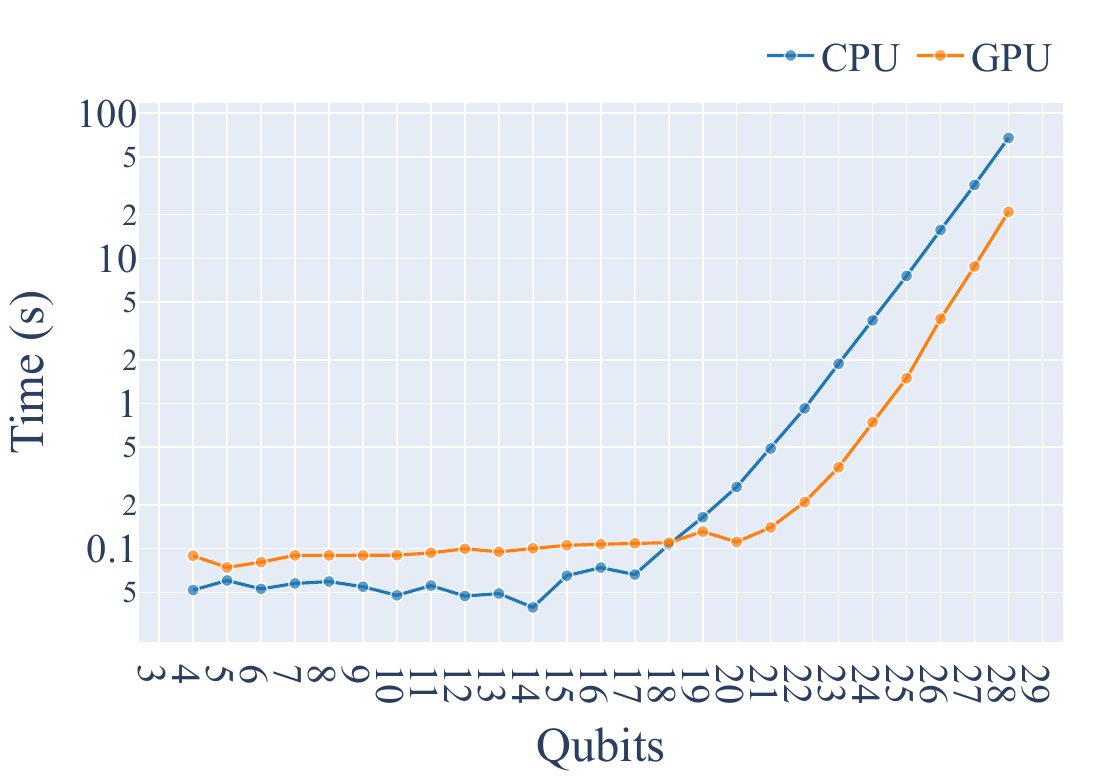}
  }\hfill
  \subfloat[Ryzen 5 5500U (Baseline).]{
    \includegraphics[width=.23\linewidth]{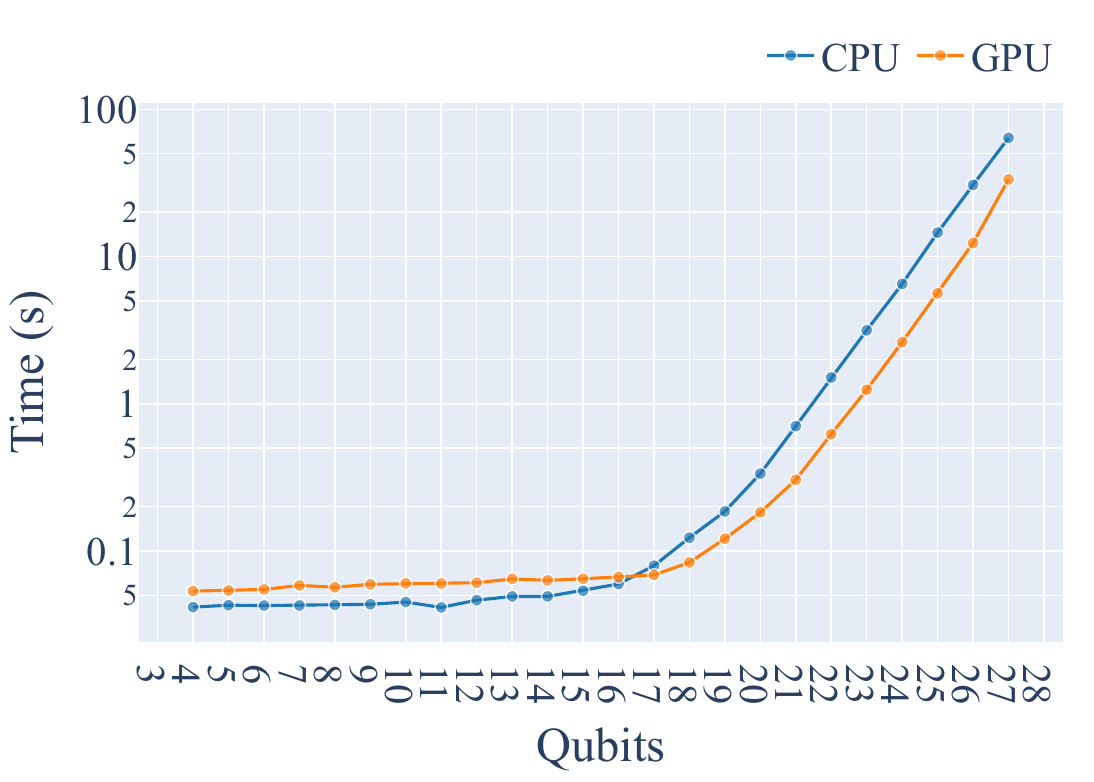}
  }\hfill
  \subfloat[M1 Pro (Baseline).]{
    \includegraphics[width=.23\linewidth]{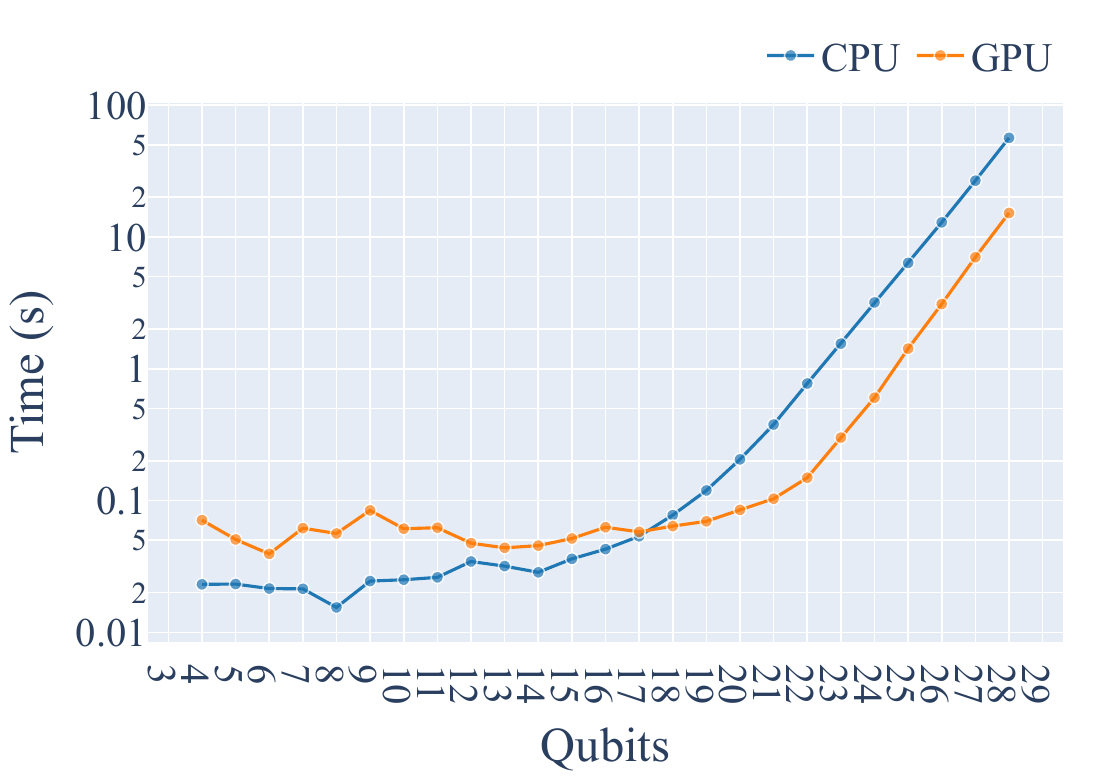}
  }

  \subfloat[Core i5-1340P (Optimized).]{
    \includegraphics[width=.23\linewidth]{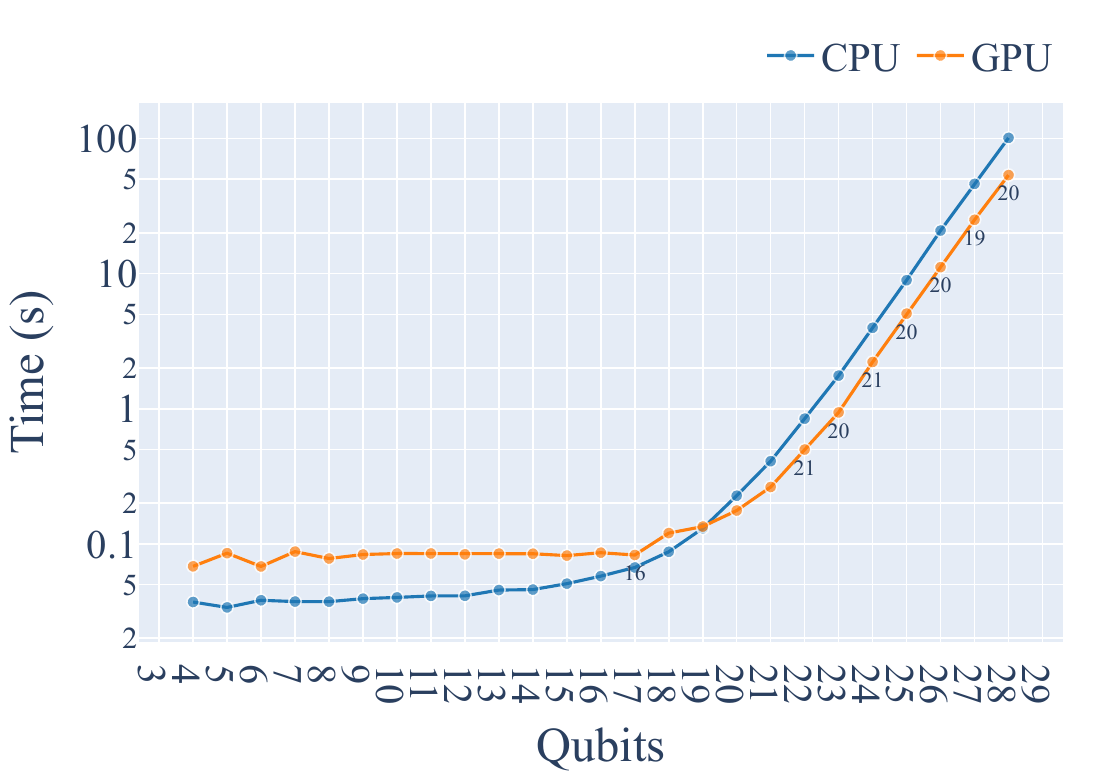}
  }\hfill
  \subfloat[Core Ultra 7 258V (Optimized).]{
    \includegraphics[width=.23\linewidth]{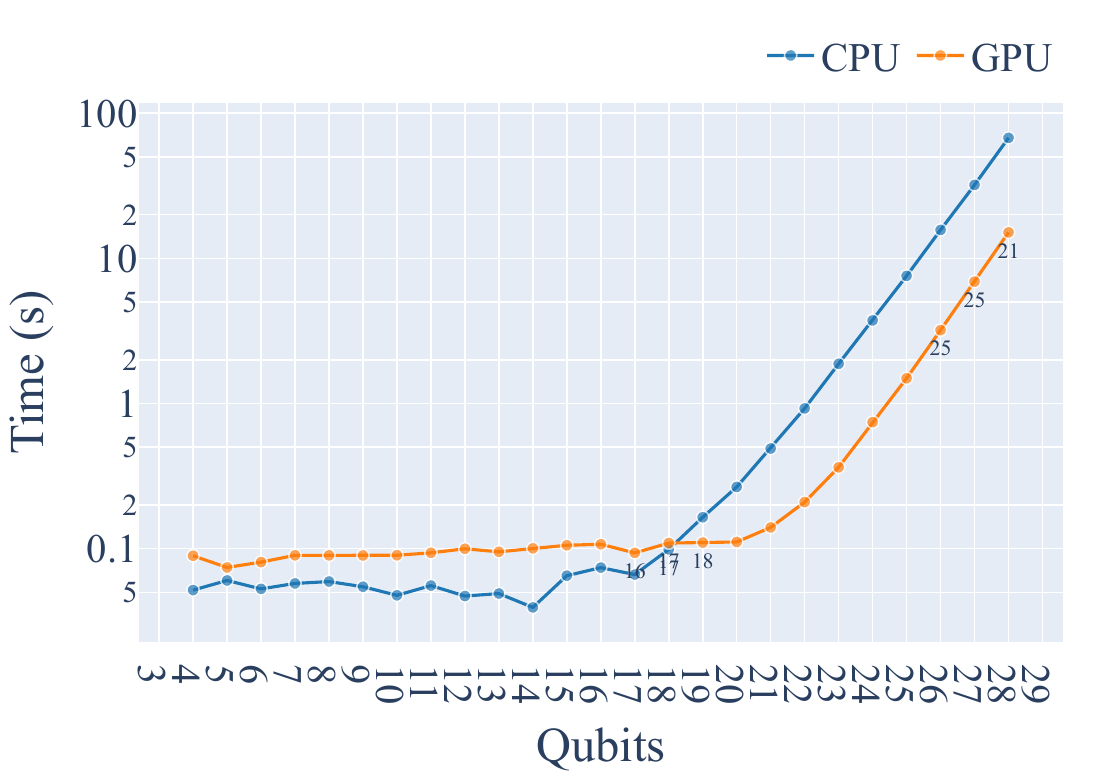}
  }\hfill
  \subfloat[Ryzen 5 5500U (Optimized).]{
    \includegraphics[width=.23\linewidth]{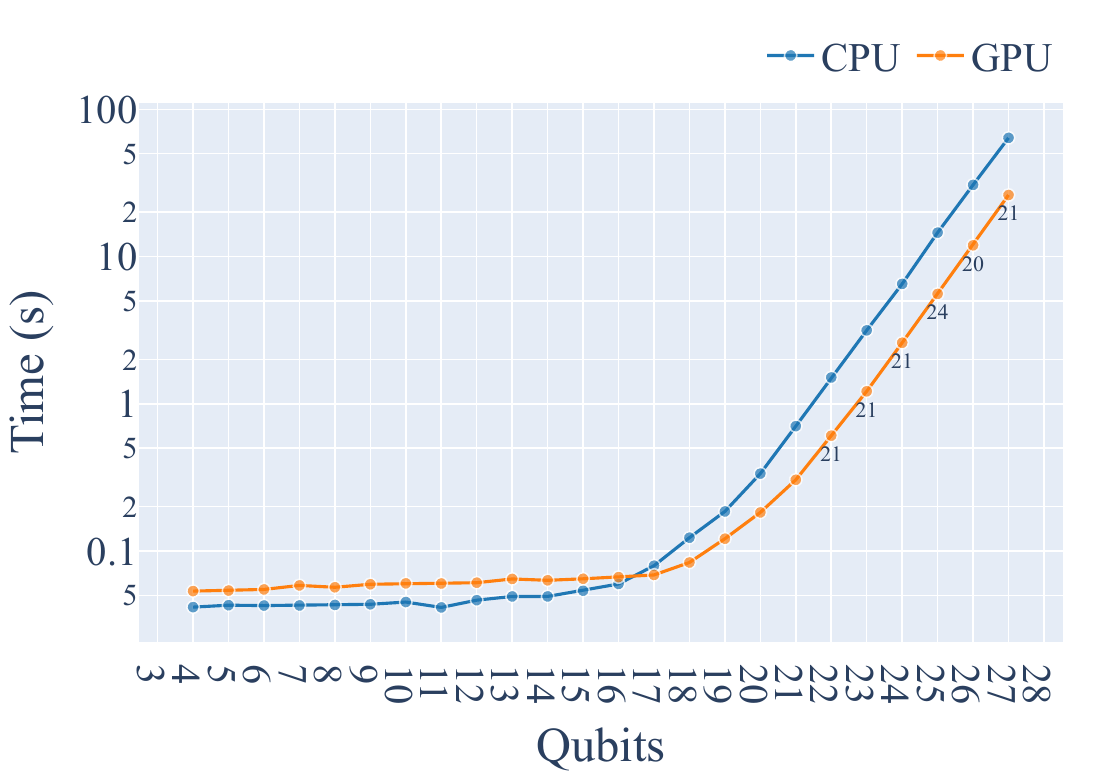}
  }\hfill
  \subfloat[M1 Pro (Optimized).]{
    \includegraphics[width=.23\linewidth]{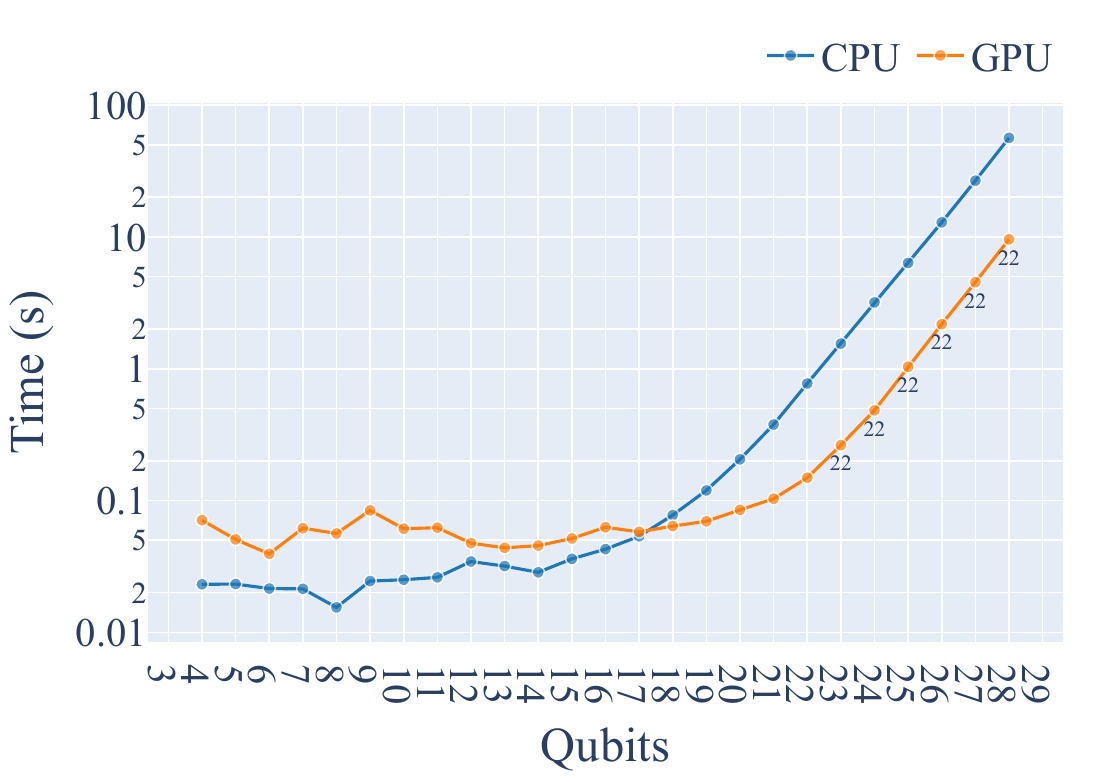}
  }

  \caption{Execution times for the QPE simulation on CPU versus integrated GPU. The top row (a--d) shows the baseline, while the bottom row (e--h) displays the execution times using the optimal state partitioning. The y-axis represents the wall-clock time in seconds on a logarithmic scale, and the x-axis indicates the number of simulated qubits. Executions range from 4 to 28 qubits, except for the AMD system (c, g), which is limited to a maximum of 27 qubits. In the optimized plots (e--h), the numbers annotated below the data points indicate the optimal block size yielding the shortest execution time.}
  \label{fig:execution_comparison}
\end{figure*}

\begin{figure*}
  \centering

  \subfloat[Intel Core i5-1340P (CPU).]{
    \includegraphics[width=.23\linewidth]{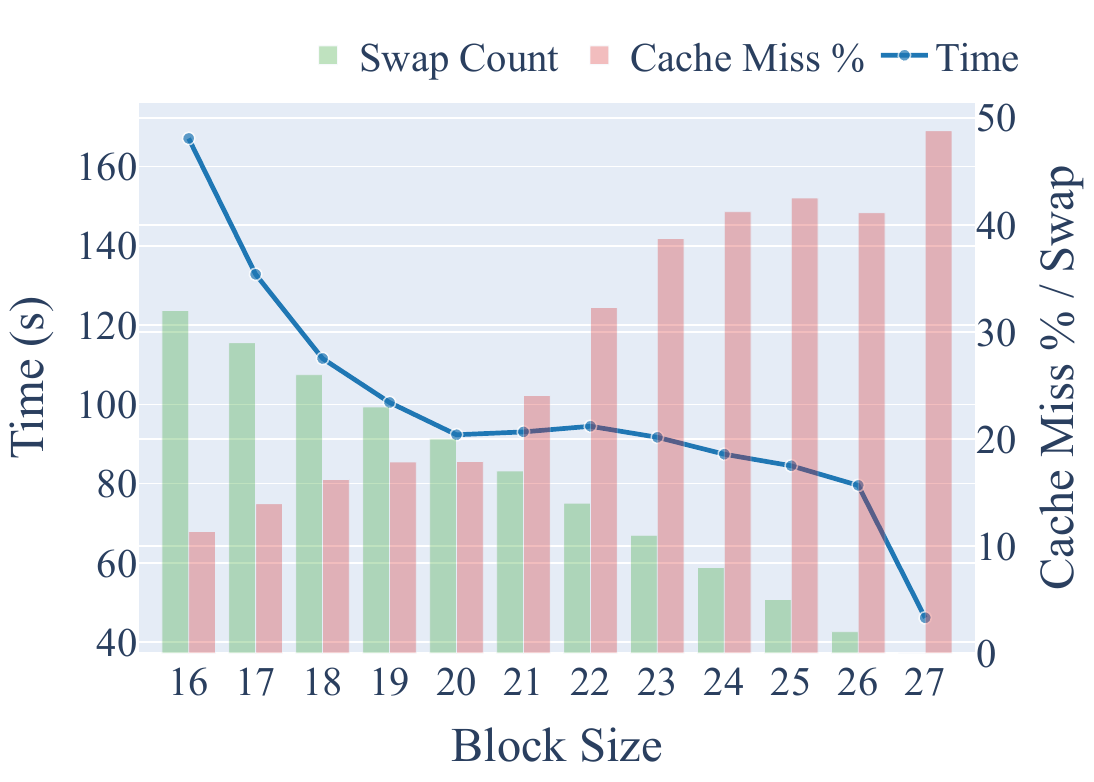}
  }
  \subfloat[Intel Core Ultra 7 258V (CPU).]{
    \includegraphics[width=.23\linewidth]{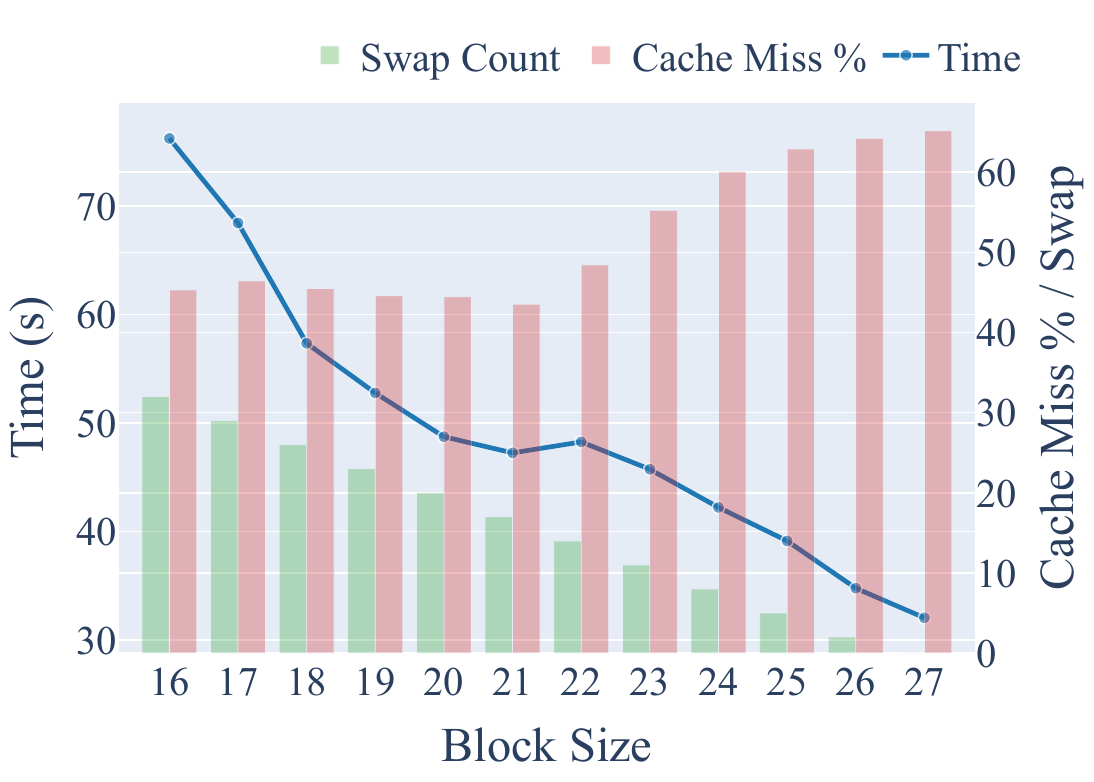}
  }
  \subfloat[AMD Ryzen 5 5500U (CPU).]{
    \includegraphics[width=.23\linewidth]{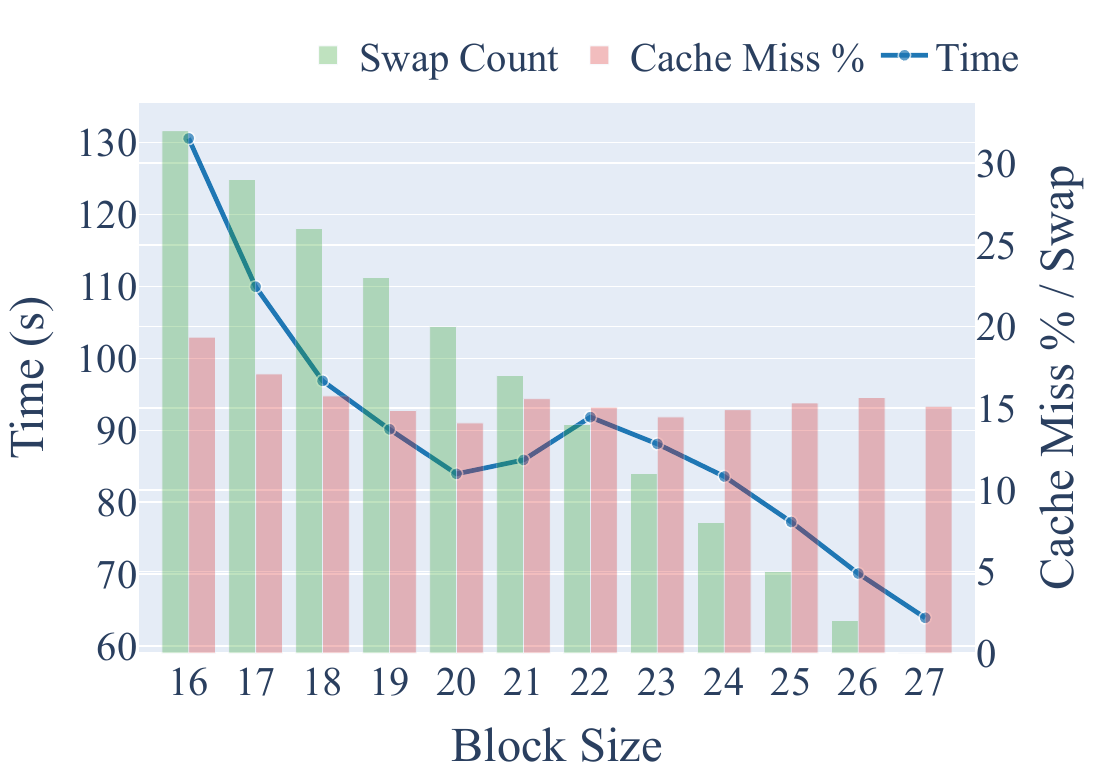}
  }
  \subfloat[Apple M1 Pro (CPU).]{
    \includegraphics[width=.23\linewidth]{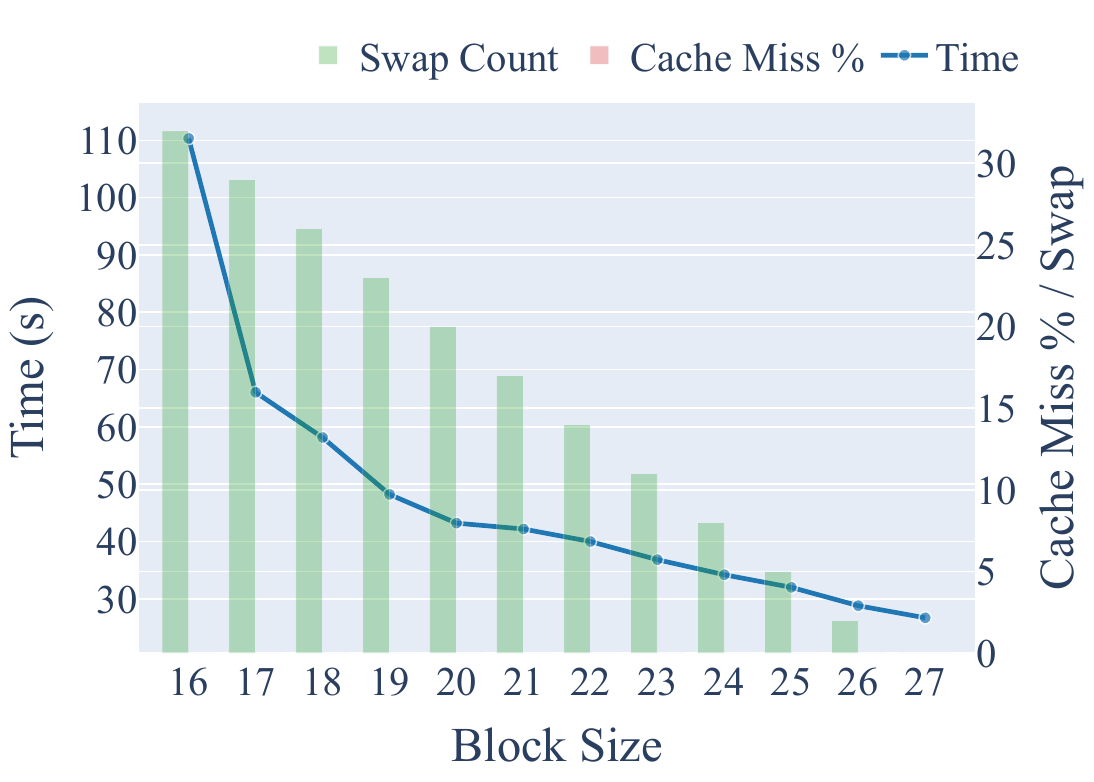}
  }

  \subfloat[Intel Core i5-1340P (GPU).]{
    \includegraphics[width=.23\linewidth]{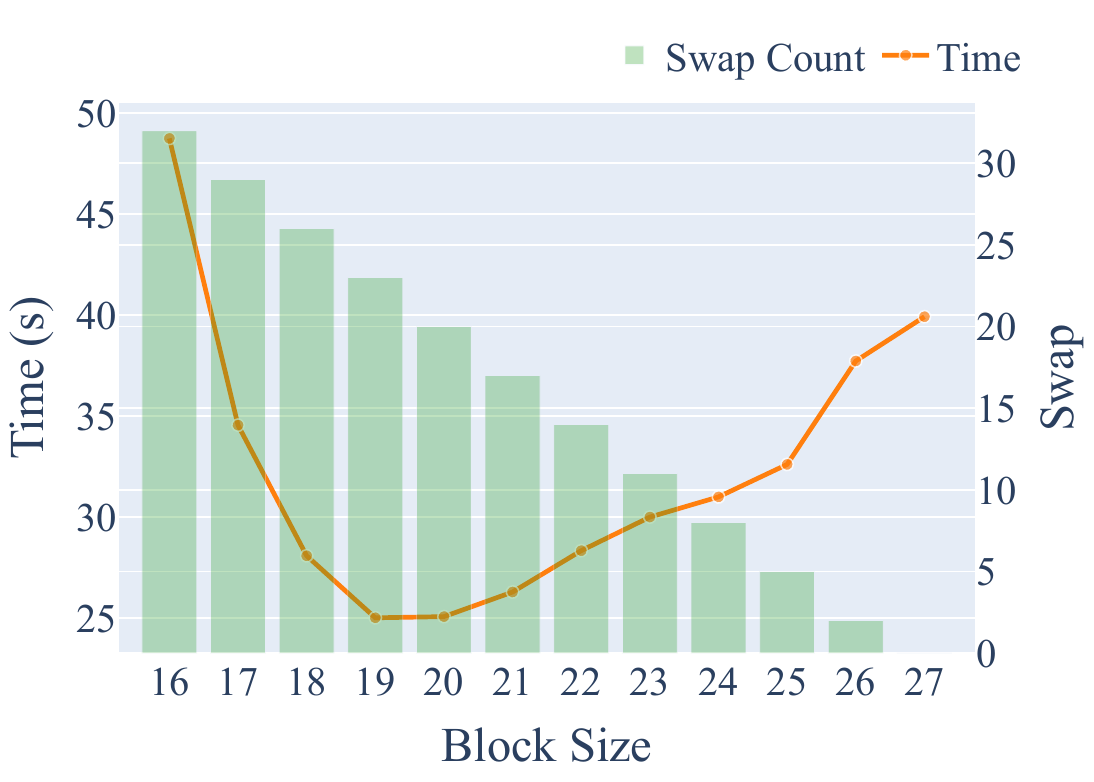}
  }
  \subfloat[Intel Core Ultra 7 258V (GPU).]{
    \includegraphics[width=.23\linewidth]{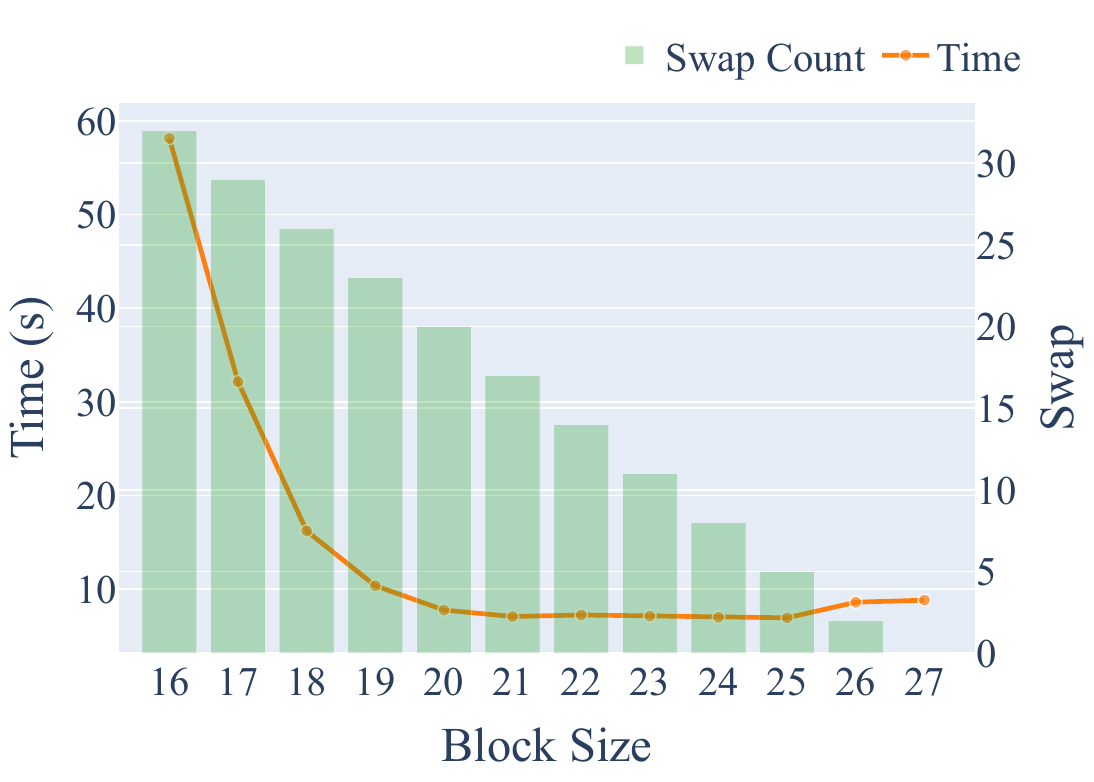}
  }
  \subfloat[AMD Ryzen 5 5500U (GPU).]{
    \includegraphics[width=.23\linewidth]{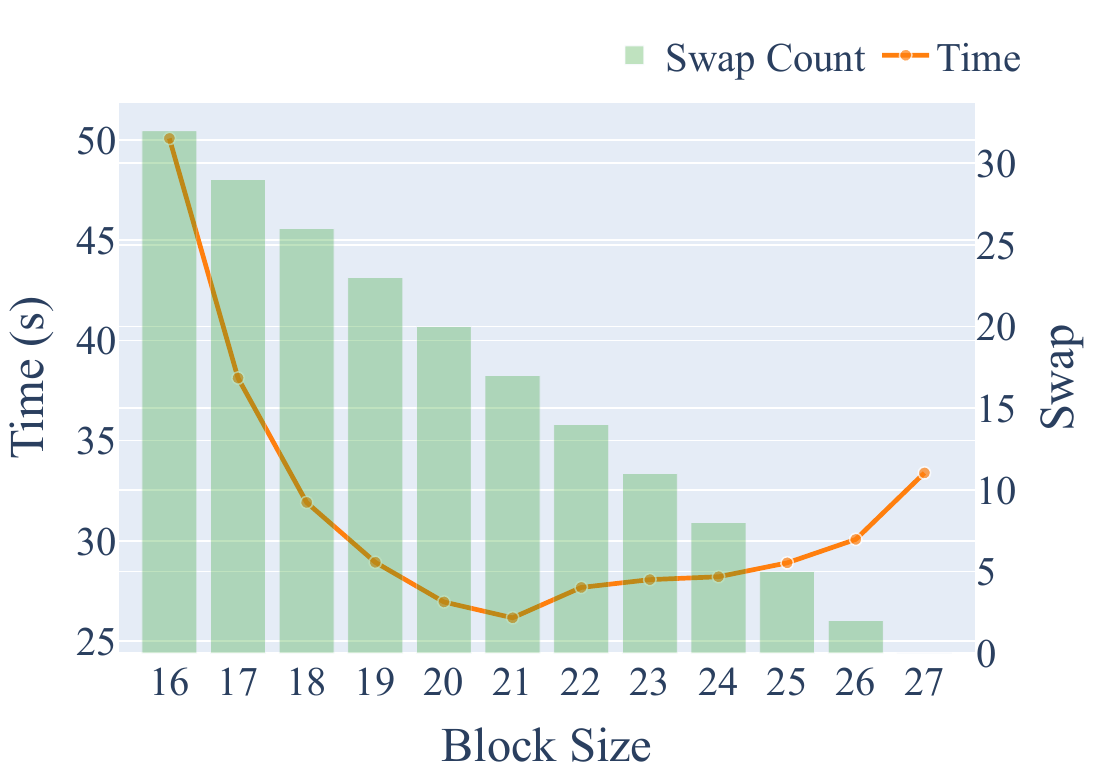}
  }
  \subfloat[Apple M1 Pro (GPU).]{
    \includegraphics[width=.23\linewidth]{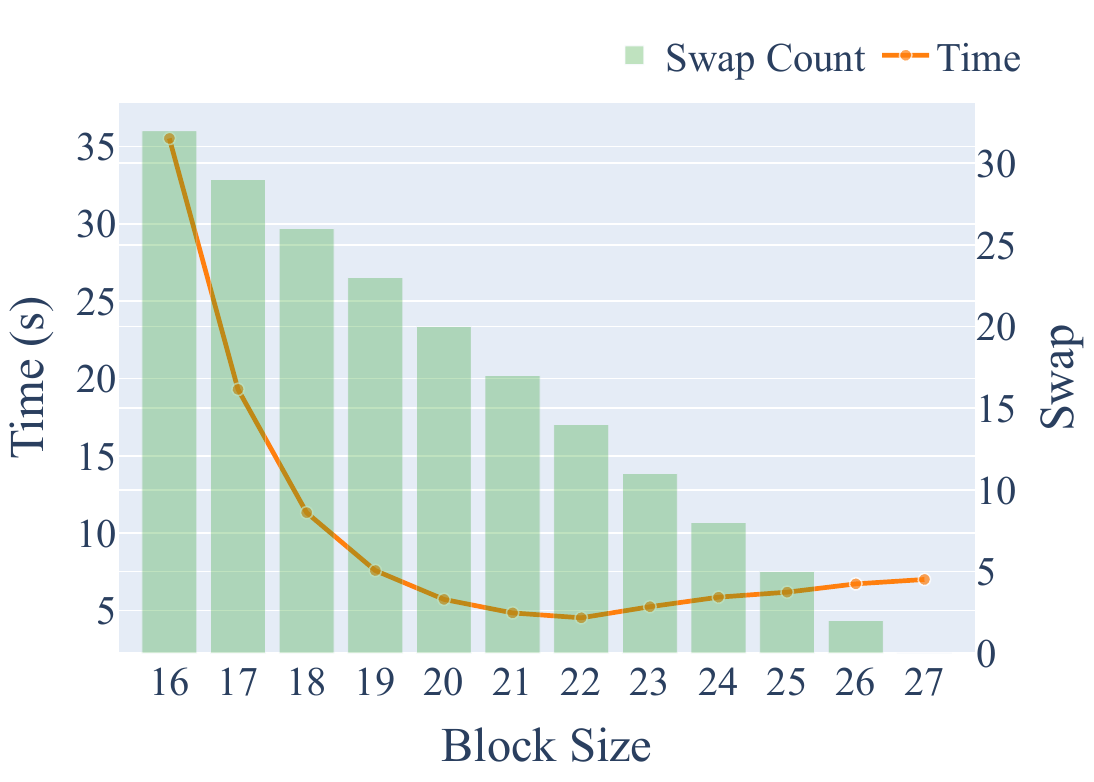}
  }

  \caption{Performance analysis of a 27-qubit QPE simulation on CPU and integrated GPU. The top row (a--d) displays the CPU execution, while the bottom row (e--h) shows the GPU execution. The plots illustrate execution time (left y-axis), cache miss percentage (red bars, right y-axis), and SWAP count (green bars, right y-axis) for block sizes ranging from 16 to 27. Cache miss data is not available GPU executions and Apple M1 Pro CPU execution (d).}  \label{fig:cache_miss}
\end{figure*}

\begin{figure*}
  \centering

  \subfloat[Intel Core i5-1340P.]{
    \includegraphics[width=.42\linewidth]{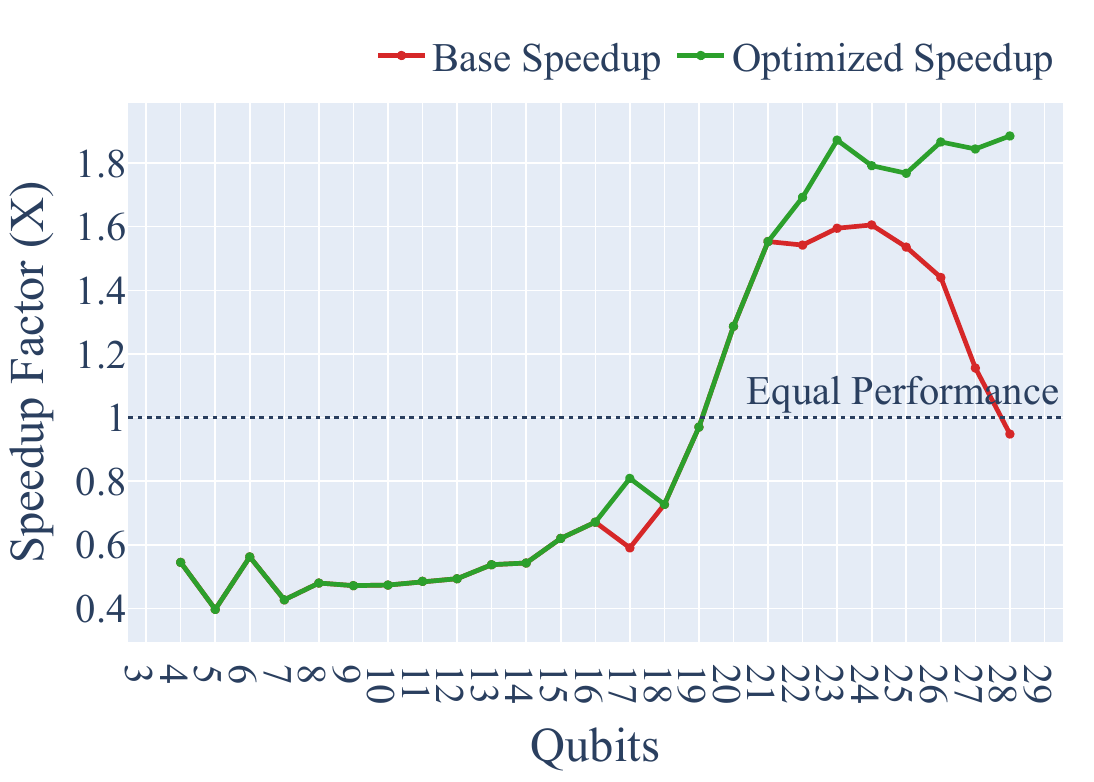}
  }\hfil
  \subfloat[Intel Core Ultra 7 258V.]{
    \includegraphics[width=.42\linewidth]{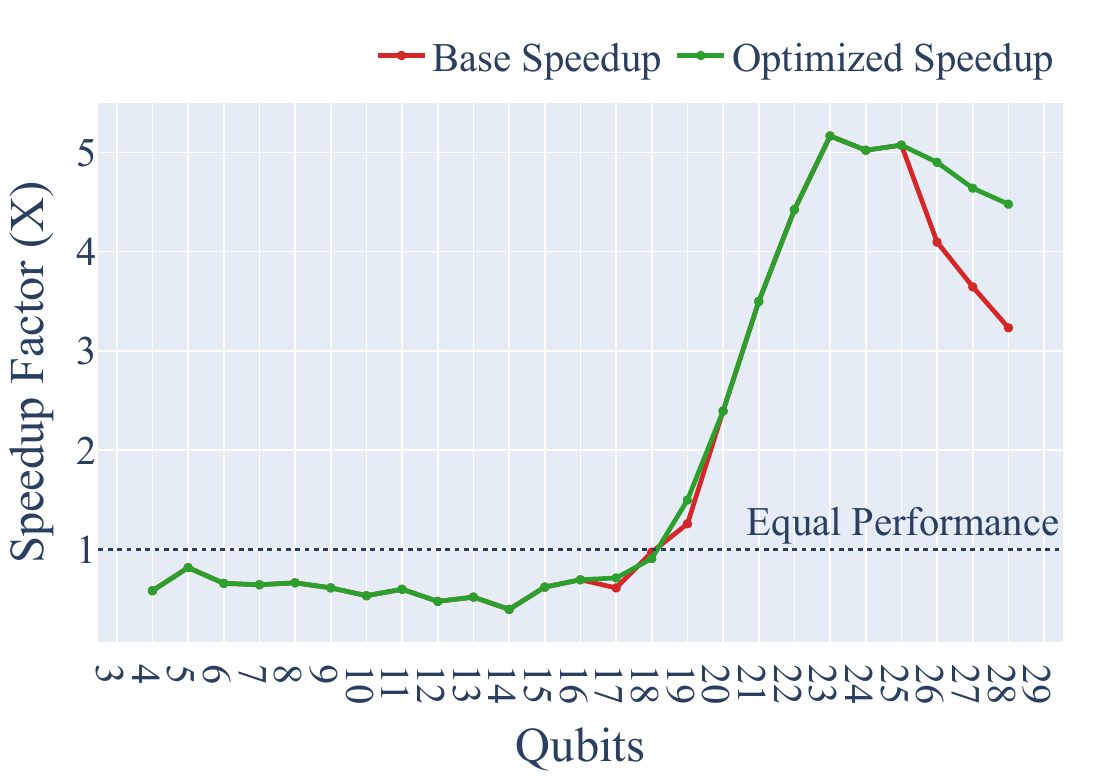}
  }

  \subfloat[AMD Ryzen 5 5500U.]{
    \includegraphics[width=.42\linewidth]{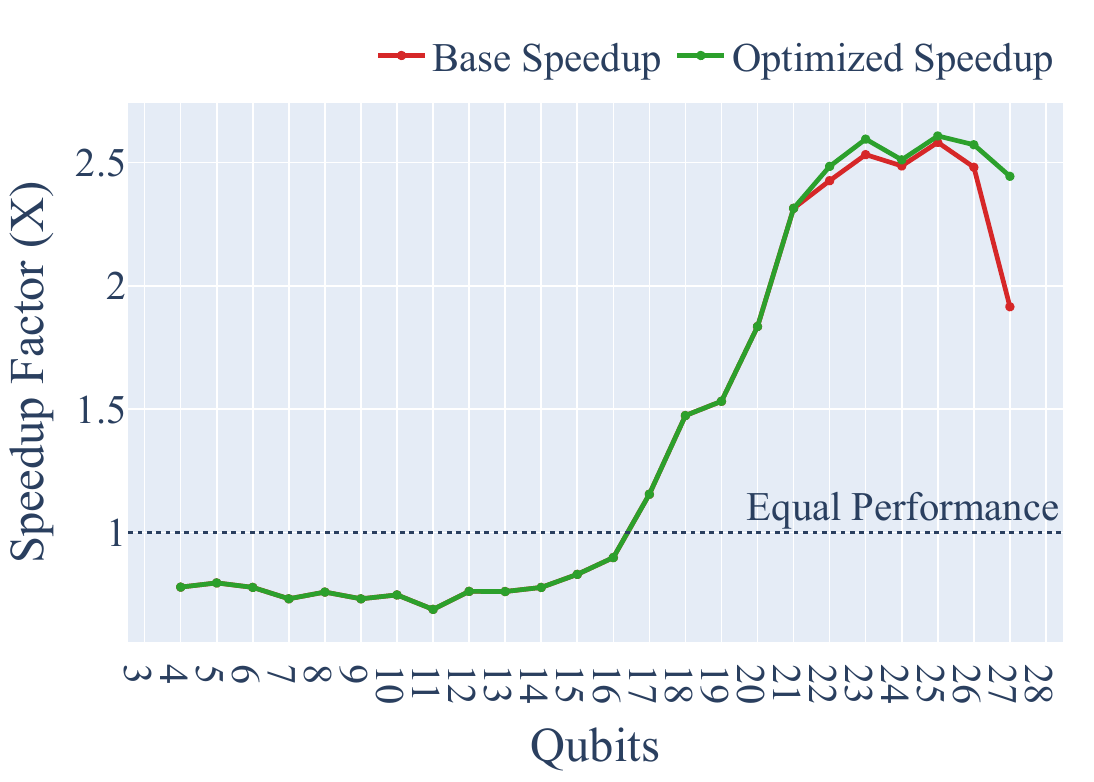}
  }\hfil
  \subfloat[Apple M1 Pro.]{
    \includegraphics[width=.42\linewidth]{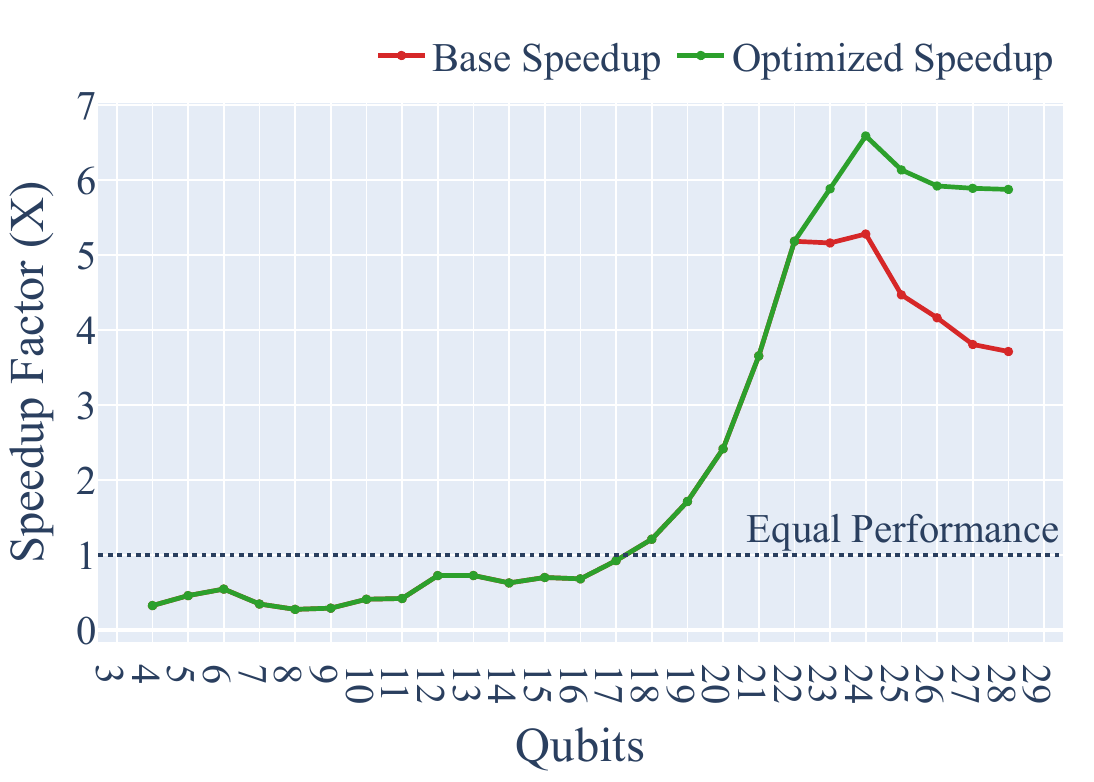}
  }

  \caption{Speedup factor of the integrated GPU relative to the CPU for the QPE simulation. The y-axis represents the relative speedup, where a value greater than 1 indicates faster execution on the GPU. The x-axis denotes the number of simulated qubits. The red line (Base Speedup) shows the relative performance without state partitioning, while the green line (Optimized Speedup) shows the performance with the proposed optimization (optimal block size).}
  \label{fig:speedup}
\end{figure*}

\section{Results and Discussion}\label{sec:results_and_discussion}

This section analyzes the impact of the proposed optimization on the four systems detailed in Table~\ref{tab:specs_notebooks}. Section~\ref{subsec:baseline} evaluates the baseline execution to establish the performance characteristics of the unoptimized simulation, providing context for the improvements achieved by the proposed state partitioning method discussed in Section~\ref{subsec:optimized}.

\subsection{Baseline}\label{subsec:baseline}

Figure~\ref{fig:execution_comparison} presents the execution times for both CPU- and GPU-accelerated simulations. In the baseline case (without state division), the expected behavior is observed across all systems: once the GPU kernel launch overhead is amortized, the GPU outperforms the CPU. This crossover typically occurs at approximately 19 qubits. Note that the experiments were performed with minimal background processes running (\textit{e.g.}, on Linux systems, the desktop environment and network manager were stopped) to reduce measurement noise. Under a fully loaded system, observations shows the GPU-accelerated simulation outperforming the CPU at smaller qubit counts.

Figure~\ref{fig:speedup} presents the speedup of the GPU-accelerated simulation relative to the CPU execution. A notable trend across all systems is a substantial decrease in the speedup factor at approximately 25 qubits. This performance degradation indicates that the simulation becomes bandwidth-bound. The memory access pattern in the state vector algorithm (Algorithm~\ref{fig:apply_gate}) is non-sequential; the inner loop reads and writes to distant memory addresses. Because every element of the $2^n$ state vector must be accessed during each quantum gate application, the algorithm exhibits poor spatial locality.

The CPU architecture is generally better equipped to handle this lack of spatial locality. Because the iterations of the inner loop are independent, CPU features such as memory-level parallelism and out-of-order execution can partially hide cache miss latency. Conversely, on a GPU, a cache miss stalls the entire thread warp executing on a Streaming Multiprocessor (SM). Due to the poor spatial locality, multiple warps may stall simultaneously while waiting for main memory accesses, thereby reducing GPU occupancy and overall throughput. In the case of the Intel Core i5-1340P, the GPU execution time exceeds that of the CPU for the 28-qubit simulation, and a general reduction in GPU speedup is observed across all evaluated systems at larger qubit counts.

\subsection{Optimized}\label{subsec:optimized}

The objective of the proposed optimization is to improve cache locality by maximizing the number of operations performed within a single state partition before transitioning to the next. This approach aims to retain the active block in the cache as long as possible. Because there are no Performance Monitoring Unit (PMU) registers readily available to directly record GPU cache misses on the evaluated integrated platforms, the L3 cache miss rate of the CPU execution is used as a proxy metric. This substitution is viable because the integrated GPU and CPU share the last-level cache (LLC).

Figure~\ref{fig:cache_miss} illustrates the cache miss rate for CPU execution of a 27-qubit QPE simulation across various block sizes. Each processor exhibits distinct cache behavior, which subsequently influences the performance gains yielded by the optimization. The results for each architecture are analyzed individually:

\subsubsection{Intel Core i5-1340P}

For the 27-qubit QPE simulation, the baseline cache miss rate is 48.80\%. A clear reduction in the miss rate is observed as the block size decreases. At a block size of 20 qubits, the point where the cache miss rate cross with the SWAP count metric, the miss rate drops to 17.89\%. It is important to note that only the miss rate of the performance cores was recorded, as these cores primarily dictate the execution time. Regarding the GPU execution time, the shortest durations are observed at block sizes of 19 and 20 qubits, measuring 25.02s and 25.08s, respectively. Assuming single-precision complex numbers (8 bytes per amplitude), a 20-qubit block requires exactly 8~MB of memory, which fits within the processor's 12~MB L3 cache. Conversely, a 21-qubit block requires 16~MB, exceeding the L3 capacity. Analyzing the optimized execution in Figure~\ref{fig:execution_comparison} confirms that the optimal block size is around 20 qubits. Furthermore, Figure~\ref{fig:speedup} shows that the proposed optimization mitigates the speedup degradation observed beyond 22 qubits. For the 28-qubit simulation, the GPU speedup recovers from 0.95$\times$ (a performance deficit) in the baseline to 1.89$\times$ with the optimization.

\subsubsection{Intel Core Ultra 7 258V}

Despite featuring a 12~MB L3 cache similar to the i5-1340P, the Core Ultra 7 258V exhibits a higher overall cache miss rate. In the baseline 27-qubit simulation, the performance cores experience a 66.18\% miss rate. The minimum recorded miss rate is 43.53\% at a block size of 21 qubits. Decreasing the block size further results in an increased miss rate, an expected outcome due to the higher frequency of required SWAP operations. Unlike the i5 processor, the optimization yields only a modest improvement in cache locality for the Ultra 7. Consequently, the reduction in GPU execution time only appears at 26 qubits and above. Nevertheless, for the 28-qubit simulation, the optimization yields a notable improvement in the GPU speedup factor, increasing from 2.23$\times$ in the baseline to 4.48$\times$. 

\subsubsection{AMD Ryzen 5 5500U}

The Ryzen 5 processor demonstrated the smallest relative impact from the proposed optimization. Interestingly, despite having the smallest L3 cache among the tested systems (8~MB), it maintained the minimum baseline cache miss rate. For the 27-qubit QPE simulation, the baseline miss rate is 16.96\%, which decreases to 14.67\% at the optimal block size of 20 qubits. As previously established, a 20-qubit block requires exactly 8~MB of memory. Corresponding to this modest reduction in cache misses, the GPU simulation exhibits a similarly modest performance gain. For the 27-qubit simulation, the optimization increases the GPU speedup factor from 1.92$\times$ to 2.44$\times$. 

\subsubsection{Apple M1 Pro}

Cache miss rates could not be recorded for the Apple M1 Pro. The M1 architecture differs fundamentally from the x86\_64 processors tested; its last-level cache, referred to as the System Level Cache (SLC), is dynamically shared among the CPU, GPU, and other coprocessors. This architecture prevents the isolation and computation of miss rates for an individual process such as the simulator. However, the M1 Pro demonstrated a consistent performance gain across circuit sizes. For simulations of 23 qubits and larger, the optimal block size was consistently identified as 22 qubits, which requires 32~MB of memory. This correlates with the hardware specifications of the processor, which features 24~MB of L2 cache for the performance cores and a 24~MB SLC. For the 28-qubits simulation, the optimization increases the GPU speedup from 3.71$\times$ to 5.88$\times$. 


While the impact varies across the evaluated systems, the proposed optimization successfully reduced execution times on all platforms. This improvement is primarily attributed to enhanced cache locality; the system that exhibited the largest reduction in cache miss rate (the i5-1340P) also demonstrated the most substantial decrease in execution time relative to its baseline. Conversely, the optimization did not yield performance gains for pure CPU executions. This result suggests that the latency incurred by cache misses on the CPU is already effectively mitigated by hardware-level memory parallelism and out-of-order execution, rendering the software-level state partitioning redundant for CPU-only simulations on these specific architectures.

\section{Final Remarks}\label{sec:conclusion}

In this paper, we presented how state division strategies, typically applied to distributed quantum simulation, can be employed to improve cache spatial locality, which positively impacts the execution time of quantum simulations on integrated GPUs. The experimental results demonstrate that, while the CPU can effectively mitigate cache miss latency via memory-level parallelism and out-of-order execution, the integrated GPU is significantly more affected by the poor spatial locality inherent to the state vector simulation algorithm.

The development of this GPU-accelerated simulator was motivated by the objective of maximizing the use of computational resources of consumer-grade laptops. Although several GPU-accelerated simulators are currently available, most rely on vendor-specific frameworks, targeting exclusively NVIDIA CUDA, AMD ROCm, or Apple Metal. This fragmentation motivated our adoption of CubeCL, a vendor-agnostic framework, that utilizes the graphics library \texttt{wgpu} to ensure broad hardware compatibility.

Although the initial implementation, which lacked state partitioning, provided measurable improvements over the CPU execution, the decline in relative speedup observed near the 27-qubit mark prompted further investigation. The hypothesis that the memory access pattern of the simulator was the primary performance bottleneck motivated the exploration of techniques to enhance cache locality. As demonstrated by our results, implementing a state partitioning strategy tailored to the processor's last-level cache successfully reduced cache miss rates. This optimization mitigated the performance degradation at larger qubit counts, yielding consistent execution time improvements across the evaluated integrated GPUs.

Despite these positive outcomes, due to hardware availability constraints, this evaluation was limited to four distinct systems. For future work, a controlled comparison of contemporaneous processors from competing vendors, specifically those from the same release cycle and with equivalent thermal design power (TDP), would be valuable to determine which architectures are most suitable for quantum simulation workloads. Such an analysis could inform the procurement decisions of quantum computing research groups when upgrading their infrastructure.

Furthermore, the current testing was restricted to integrated GPUs, and this analysis does not directly translate to discrete GPUs, which possess different memory hierarchies and bandwidth capabilities. Therefore, evaluating this state partitioning optimization on discrete GPUs from vendors such as Intel, NVIDIA, and AMD remains a subject for future research.

\section*{Acknowledgment}
This work was executed under the TIC26 -- Brazil Quantum Camp project, funded within the scope of the Prioritized Informatics Programs and Projects (PPI), Process No. 01245.008254/2025-22, under the responsibility of the Ministry of Science, Technology and Innovation (MCTI), with operational coordination by the Association for the Promotion of Brazilian Software Excellence (SOFTEX), and executed by CESAR and the Instituto de Pesquisas Eldorado.

Artificial Intelligence (AI) tools, specifically Google's Gemini 3.1 Pro, were utilized for editing and grammar enhancement during the preparation of this manuscript. The authors adopted a methodology in which all sections were initially drafted without AI assistance, followed by iterative refinement and proofreading using the model. In the software development phase, Gemini was employed to identify potential bugs within the simulator's source code. Furthermore, the scripts used to automate the benchmarks and generate the plots were initially drafted by the AI, followed by comprehensive human validation and verification.

\bibliographystyle{IEEEtran}
\bibliography{main}

@inproceedings{bayraktar2023,
  title = {{{cuQuantum SDK}}: {{A High-Performance Library}} for {{Accelerating Quantum Science}}},
  shorttitle = {{{cuQuantum SDK}}},
  booktitle = {2023 {{IEEE International Conference}} on {{Quantum Computing}} and {{Engineering}} ({{QCE}})},
  author = {Bayraktar, Harun and Charara, Ali and Clark, David and Cohen, Saul and Costa, Timothy and Fang, Yao-Lung L. and Gao, Yang and Guan, Jack and Gunnels, John and Haidar, Azzam and Hehn, Andreas and Hohnerbach, Markus and Jones, Matthew and Lubowe, Tom and Lyakh, Dmitry and Morino, Shinya and Springer, Paul and Stanwyck, Sam and Terentyev, Igor and Varadhan, Satya and Wong, Jonathan and Yamaguchi, Takuma},
  year = 2023,
  month = sep,
  pages = {1050--1061},
  publisher = {IEEE},
  address = {Bellevue, WA, USA},
  doi = {10.1109/QCE57702.2023.00119},
  urldate = {2026-02-11},
  copyright = {https://doi.org/10.15223/policy-029},
  isbn = {979-8-3503-4323-6}
}

@article{cicero2026,
  title = {Simulation of {{Quantum Computers}}: {{Review}} and {{Acceleration Opportunities}}},
  shorttitle = {Simulation of {{Quantum Computers}}},
  author = {Cicero, Alessio and Maleki, Mohammad Ali and Azhar, Muhammad Waqar and Kockum, Anton Frisk and Trancoso, Pedro},
  year = 2026,
  month = mar,
  journal = {ACM Trans. Quantum Comput.},
  volume = {7},
  number = {1},
  pages = {1--35},
  issn = {2643-6809, 2643-6817},
  doi = {10.1145/3762672},
  urldate = {2026-02-23},
  langid = {english}
}

@inproceedings{doi2020,
  title = {Cache {{Blocking Technique}} to {{Large Scale Quantum Computing Simulation}} on {{Supercomputers}}},
  booktitle = {2020 {{IEEE International Conference}} on {{Quantum Computing}} and {{Engineering}} ({{QCE}})},
  author = {Doi, Jun and Horii, Hiroshi},
  year = 2020,
  month = oct,
  pages = {212--222},
  publisher = {IEEE},
  address = {Denver, CO, USA},
  doi = {10.1109/QCE49297.2020.00035},
  urldate = {2026-02-24},
  copyright = {https://ieeexplore.ieee.org/Xplorehelp/downloads/license-information/IEEE.html},
  isbn = {978-1-7281-8969-7}
}

@inproceedings{fang2022,
  title = {Efficient {{Hierarchical State Vector Simulation}} of {{Quantum Circuits}} via {{Acyclic Graph Partitioning}}},
  booktitle = {2022 {{IEEE International Conference}} on {{Cluster Computing}} ({{CLUSTER}})},
  author = {Fang, Bo and Ozkaya, M. Yusuf and Li, Ang and Catalyurek, Umit V. and Krishnamoorthy, Sriram},
  year = 2022,
  month = sep,
  pages = {289--300},
  publisher = {IEEE},
  address = {Heidelberg, Germany},
  doi = {10.1109/CLUSTER51413.2022.00041},
  urldate = {2026-02-23},
  copyright = {https://doi.org/10.15223/policy-029},
  isbn = {978-1-6654-9856-2}
}

@misc{gottesman1997,
  title = {Stabilizer {{Codes}} and {{Quantum Error Correction}}},
  author = {Gottesman, Daniel},
  year = 1997,
  month = may,
  number = {arXiv:quant-ph/9705052},
  eprint = {quant-ph/9705052},
  publisher = {arXiv},
  doi = {10.48550/arXiv.quant-ph/9705052},
  urldate = {2026-02-23},
  archiveprefix = {arXiv}
}

@article{guerreschi2020,
  title = {Intel {{Quantum Simulator}}: A Cloud-Ready High-Performance Simulator of Quantum Circuits},
  shorttitle = {Intel {{Quantum Simulator}}},
  author = {Guerreschi, Gian Giacomo and Hogaboam, Justin and Baruffa, Fabio and Sawaya, Nicolas P D},
  year = 2020,
  month = jul,
  journal = {Quantum Sci. Technol.},
  volume = {5},
  number = {3},
  pages = {034007},
  issn = {2058-9565},
  doi = {10.1088/2058-9565/ab8505},
  urldate = {2026-02-11}
}

@article{jones2019,
  title = {{{QuEST}} and {{High Performance Simulation}} of {{Quantum Computers}}},
  author = {Jones, Tyson and Brown, Anna and Bush, Ian and Benjamin, Simon C.},
  year = 2019,
  month = jul,
  journal = {Sci Rep},
  volume = {9},
  number = {1},
  pages = {10736},
  issn = {2045-2322},
  doi = {10.1038/s41598-019-47174-9},
  urldate = {2026-02-11},
  langid = {english}
}

@inproceedings{li2021,
  title = {{{SV-sim}}: Scalable {{PGAS-based}} State Vector Simulation of Quantum Circuits},
  shorttitle = {{{SV-sim}}},
  booktitle = {Proceedings of the {{International Conference}} for {{High Performance Computing}}, {{Networking}}, {{Storage}} and {{Analysis}}},
  author = {Li, Ang and Fang, Bo and Granade, Christopher and Prawiroatmodjo, Guen and Heim, Bettina and Roetteler, Martin and Krishnamoorthy, Sriram},
  year = 2021,
  month = nov,
  pages = {1--14},
  publisher = {ACM},
  address = {St. Louis Missouri},
  doi = {10.1145/3458817.3476169},
  urldate = {2026-02-19},
  isbn = {978-1-4503-8442-1},
  langid = {english}
}

@article{murillo2025,
  title = {Quantum {{Software Engineering}}: {{Roadmap}} and {{Challenges Ahead}}},
  shorttitle = {Quantum {{Software Engineering}}},
  author = {Murillo, Juan Manuel and {Garcia-Alonso}, Jose and Moguel, Enrique and Barzen, Johanna and Leymann, Frank and Ali, Shaukat and Yue, Tao and Arcaini, Paolo and {P{\'e}rez-Castillo}, Ricardo and {Garc{\'i}a-Rodr{\'i}guez De Guzm{\'a}n}, Ignacio and Piattini, Mario and {Ruiz-Cort{\'e}s}, Antonio and Brogi, Antonio and Zhao, Jianjun and Miranskyy, Andriy and Wimmer, Manuel},
  year = 2025,
  month = jun,
  journal = {ACM Trans. Softw. Eng. Methodol.},
  volume = {34},
  number = {5},
  pages = {1--48},
  issn = {1049-331X, 1557-7392},
  doi = {10.1145/3712002},
  urldate = {2026-02-23},
  langid = {english}
}

@book{nielsen2010,
  title = {Quantum Computation and Quantum Information},
  author = {Nielsen, Michael A. and Chuang, Isaac L.},
  year = 2010,
  edition = {10th anniversary edition},
  publisher = {Cambridge university press},
  address = {Cambridge},
  doi = {10.1017/CBO9780511976667},
  isbn = {978-1-107-00217-3},
  langid = {english},
  lccn = {530.12}
}

@article{preskill2018,
  title = {Quantum {{Computing}} in the {{NISQ}} Era and Beyond},
  author = {Preskill, John},
  year = 2018,
  month = aug,
  journal = {Quantum},
  volume = {2},
  pages = {79},
  issn = {2521-327X},
  doi = {10.22331/q-2018-08-06-79},
  urldate = {2026-02-23},
  copyright = {https://creativecommons.org/licenses/by/4.0/},
  langid = {english}
}

@article{rosa2026,
  title = {Full {{Quantum Stack}}: {{Ket Platform}}},
  shorttitle = {Full {{Quantum Stack}}},
  author = {Rosa, Evandro and Lussi, Eduardo and Marchi, Jerusa and De Santiago, Rafael and Duzzioni, Eduardo},
  year = 2026,
  month = feb,
  journal = {Braz J Phys},
  volume = {56},
  number = {1},
  pages = {45},
  issn = {0103-9733, 1678-4448},
  doi = {10.1007/s13538-025-01981-w},
  urldate = {2026-02-11},
  langid = {english}
}

@misc{smelyanskiy2016,
  title = {{{qHiPSTER}}: {{The Quantum High Performance Software Testing Environment}}},
  shorttitle = {{{qHiPSTER}}},
  author = {Smelyanskiy, Mikhail and Sawaya, Nicolas P. D. and {Aspuru-Guzik}, Al{\'a}n},
  year = 2016,
  month = may,
  number = {arXiv:1601.07195},
  eprint = {1601.07195},
  primaryclass = {quant-ph},
  publisher = {arXiv},
  doi = {10.48550/arXiv.1601.07195},
  urldate = {2026-02-11},
  archiveprefix = {arXiv}
}

@inproceedings{zhang2021,
  title = {{{HyQuas}}: Hybrid Partitioner Based Quantum Circuit Simulation System on {{GPU}}},
  shorttitle = {{{HyQuas}}},
  booktitle = {Proceedings of the {{ACM International Conference}} on {{Supercomputing}}},
  author = {Zhang, Chen and Song, Zeyu and Wang, Haojie and Rong, Kaiyuan and Zhai, Jidong},
  year = 2021,
  month = jun,
  pages = {443--454},
  publisher = {ACM},
  address = {Virtual Event USA},
  doi = {10.1145/3447818.3460357},
  urldate = {2026-02-23},
  isbn = {978-1-4503-8335-6},
  langid = {english}
}

\end{document}